%% file: main.tex
\documentclass[journal=jctcce,manuscript=article]{achemso}
\setkeys{acs}{articletitle = true, chaptertitle = true}

\usepackage[version=3]{mhchem}
\usepackage{atbegshi}
\usepackage{amsmath}
\usepackage{graphicx}
\usepackage{booktabs}
\usepackage{multirow}
\usepackage{siunitx}
\input{authors}

\input{title}
\input{figures/results-gpu-values}

\begin{document}

\begin{abstract}
  \input{sections/abstract}
\end{abstract}

\input{sections/intro}
\input{sections/methods}
\input{sections/protocol}
\input{sections/results}
\input{sections/conclusion}
\input{sections/availability}
\bibliography{combined}

\end{document}

%% file: authors.tex
\author{Shengrui Xu}
\altaffiliation{These authors contributed equally to this work.}
\affiliation{School of Artificial Intelligence and Data Science,
University of Science and Technology of China,
Hefei 230026,
P.R.~China}
\alsoaffiliation{SII-ECNU Joint Autonomous Chemistry Laboratory,
Shanghai Innovation Institute,
Shanghai 200003,
P.R. China}
\alsoaffiliation{Suzhou Institute for Advanced Research,
University of Science and Technology of China,
Suzhou 215123,
P.R.~China}

\author{Longhan Shen}
\altaffiliation{These authors contributed equally to this work.}
\affiliation{School of Artificial Intelligence and Data Science,
University of Science and Technology of China,
Hefei 230026,
P.R.~China}
\alsoaffiliation{Suzhou Institute for Advanced Research,
University of Science and Technology of China,
Suzhou 215123,
P.R.~China}

\author{Wenxi Zhai}
\affiliation{School of Artificial Intelligence and Data Science,
University of Science and Technology of China,
Hefei 230026,
P.R.~China}
\alsoaffiliation{SII-ECNU Joint Autonomous Chemistry Laboratory,
Shanghai Innovation Institute,
Shanghai 200003,
P.R. China}
\alsoaffiliation{Suzhou Institute for Advanced Research,
University of Science and Technology of China,
Suzhou 215123,
P.R.~China}

\author{Tong Zhu}
\affiliation{Shanghai Engineering Research Center of Molecular Therapeutics and New Drug Development,
School of Chemistry and Molecular Engineering,
East China Normal University,
Shanghai 200062,
P.R.~China}
\alsoaffiliation{SII-ECNU Joint Autonomous Chemistry Laboratory,
Shanghai Innovation Institute,
Shanghai 200003,
P.R. China}
\email{tzhu@lps.ecnu.edu.cn}

\author{Jinzhe Zeng}
\affiliation{School of Artificial Intelligence and Data Science,
University of Science and Technology of China,
Hefei 230026,
P.R.~China}
\alsoaffiliation{State Key Laboratory of Precision and Intelligent Chemistry,
University of Science and Technology of China,
Hefei 230026,
P.R.~China}
\alsoaffiliation{Suzhou Institute for Advanced Research,
University of Science and Technology of China,
Suzhou 215123,
P.R.~China}
\alsoaffiliation{Suzhou Big Data \& AI Research and Engineering Center,
Suzhou 215123,
P.R.~China}
\email{jinzhe.zeng@ustc.edu.cn}

%% file: title.tex
\title{GenerativeQC: A Generative, GPU-Accelerated Quantum Chemistry Program}

%% file: figures/results-gpu-values.tex
\providecommand{\GQC}[1]{\ifcsname GQC@#1\endcsname\else\errmessage{Undefined GQC value #1}\fi\csname GQC@#1\endcsname}
\expandafter\def\csname GQC@cbs.alkane.direction.bound\endcsname{2.4\times10^{-8}}
\expandafter\def\csname GQC@cbs.alkane.energy.bound\endcsname{2.4\times10^{-9}}
\expandafter\def\csname GQC@cbs.all.energy.bound\endcsname{2.1\times10^{-8}}
\expandafter\def\csname GQC@cbs.all.force.bound\endcsname{1.2\times10^{-7}}
\expandafter\def\csname GQC@cbs.ctwo.direction.bound\endcsname{1.1\times10^{-8}}
\expandafter\def\csname GQC@cbs.ctwo.energy.bound\endcsname{6.3\times10^{-10}}
\expandafter\def\csname GQC@cbs.h100.first\endcsname{1.1}
\expandafter\def\csname GQC@cbs.h100.last\endcsname{267}
\expandafter\def\csname GQC@cbs.h100.last.min\endcsname{4.5}
\expandafter\def\csname GQC@cbs.h100.saving.last\endcsname{14}
\expandafter\def\csname GQC@cbs.h100.share.first\endcsname{46}
\expandafter\def\csname GQC@cbs.h100.share.last\endcsname{90}
\expandafter\def\csname GQC@cbs.h100.spread\endcsname{1}
\expandafter\def\csname GQC@cbs.h100.tlambda.last\endcsname{68}
\expandafter\def\csname GQC@cbs.h100.val.energy\endcsname{1.46\times10^{-8}}
\expandafter\def\csname GQC@cbs.h100.val.force\endcsname{4.85\times10^{-8}}
\expandafter\def\csname GQC@cbs.rtx5090.first\endcsname{3.2}
\expandafter\def\csname GQC@cbs.rtx5090.last\endcsname{1019}
\expandafter\def\csname GQC@cbs.rtx5090.last.min\endcsname{17}
\expandafter\def\csname GQC@cbs.rtx5090.saving.last\endcsname{15}
\expandafter\def\csname GQC@cbs.rtx5090.share.first\endcsname{38}
\expandafter\def\csname GQC@cbs.rtx5090.share.last\endcsname{94}
\expandafter\def\csname GQC@cbs.rtx5090.spread\endcsname{2}
\expandafter\def\csname GQC@cbs.rtx5090.tlambda.last\endcsname{78}
\expandafter\def\csname GQC@cbs.rtx5090.val.energy\endcsname{2.09\times10^{-8}}
\expandafter\def\csname GQC@cbs.rtx5090.val.force\endcsname{1.13\times10^{-7}}
\expandafter\def\csname GQC@cbs.strict.energy.bound\endcsname{2.4\times10^{-9}}
\expandafter\def\csname GQC@cbs.strict.force.bound\endcsname{1.3\times10^{-9}}
\expandafter\def\csname GQC@cbs.val13.energy.bound\endcsname{2.1\times10^{-8}}
\expandafter\def\csname GQC@cbs.val13.force.bound\endcsname{1.2\times10^{-7}}
\expandafter\def\csname GQC@hfdft.directdiff.energy\endcsname{1.6\times10^{-9}}
\expandafter\def\csname GQC@hfdft.directdiff.force\endcsname{3.7\times10^{-6}}
\expandafter\def\csname GQC@hfdft.directdiff.frms\endcsname{2.0\times10^{-6}}
\expandafter\def\csname GQC@hfdft.errdiff.energy\endcsname{1.6\times10^{-9}}
\expandafter\def\csname GQC@hfdft.errdiff.force\endcsname{3.5\times10^{-6}}
\expandafter\def\csname GQC@hfdft.error.dfhf.force\endcsname{2.0\times10^{-5}}
\expandafter\def\csname GQC@hfdft.error.dfhf.max\endcsname{2.11\times10^{-5}}
\expandafter\def\csname GQC@hfdft.error.dfhf.min\endcsname{1.85\times10^{-5}}
\expandafter\def\csname GQC@hfdft.error.directdft.force\endcsname{5.5\times10^{-5}}
\expandafter\def\csname GQC@hfdft.error.directhf\endcsname{8.9\times10^{-12}}
\expandafter\def\csname GQC@hfdft.error.directhf.force\endcsname{9.8\times10^{-7}}
\expandafter\def\csname GQC@hfdft.error.energy\endcsname{2.1\times10^{-5}}
\expandafter\def\csname GQC@hfdft.error.fmax\endcsname{5.7\times10^{-5}}
\expandafter\def\csname GQC@hfdft.gpudiff.energy\endcsname{1.9\times10^{-11}}
\expandafter\def\csname GQC@hfdft.gpudiff.force\endcsname{1.7\times10^{-6}}
\expandafter\def\csname GQC@hfdft.gqc.max.energy\endcsname{2.11\times10^{-5}}
\expandafter\def\csname GQC@hfdft.gqc.max.fmax\endcsname{5.69\times10^{-5}}
\expandafter\def\csname GQC@hfdft.gqc.max.frms\endcsname{3.32\times10^{-5}}
\expandafter\def\csname GQC@hfdft.h100.both.wins\endcsname{64}
\expandafter\def\csname GQC@hfdft.h100.cold.median.3\endcsname{6.35}
\expandafter\def\csname GQC@hfdft.h100.cold.median.96\endcsname{1.79}
\expandafter\def\csname GQC@hfdft.h100.cold.wins\endcsname{64}
\expandafter\def\csname GQC@hfdft.h100.faster\endcsname{in all 128 comparisons}
\expandafter\def\csname GQC@hfdft.h100.faster.sized\endcsname{in every cold and warm comparison}
\expandafter\def\csname GQC@hfdft.h100.large.warm.gqc.max\endcsname{4.4}
\expandafter\def\csname GQC@hfdft.h100.large.warm.gqc.min\endcsname{0.64}
\expandafter\def\csname GQC@hfdft.h100.saved.warm.3\endcsname{0.48}
\expandafter\def\csname GQC@hfdft.h100.saved.warm.96\endcsname{1.1}
\expandafter\def\csname GQC@hfdft.h100.small.warm.gpu4pyscf.ms\endcsname{550}
\expandafter\def\csname GQC@hfdft.h100.small.warm.gqc.ms\endcsname{32}
\expandafter\def\csname GQC@hfdft.h100.total.cold.gpu4pyscf\endcsname{555}
\expandafter\def\csname GQC@hfdft.h100.total.cold.gqc\endcsname{235}
\expandafter\def\csname GQC@hfdft.h100.total.warm.gpu4pyscf\endcsname{113}
\expandafter\def\csname GQC@hfdft.h100.total.warm.gqc\endcsname{43}
\expandafter\def\csname GQC@hfdft.h100.warm.median.3\endcsname{18.3}
\expandafter\def\csname GQC@hfdft.h100.warm.median.96\endcsname{1.80}
\expandafter\def\csname GQC@hfdft.h100.warm.wins\endcsname{64}
\expandafter\def\csname GQC@hfdft.h100.wbmv.max\endcsname{3.0}
\expandafter\def\csname GQC@hfdft.h100.wbmv.min\endcsname{2.2}
\expandafter\def\csname GQC@hfdft.h100.wbmv96.gpu4pyscf\endcsname{12.6}
\expandafter\def\csname GQC@hfdft.h100.wbmv96.gqc\endcsname{4.27}
\expandafter\def\csname GQC@hfdft.h100.wbmv96.path\endcsname{Direct}
\expandafter\def\csname GQC@hfdft.ref.cpu.energy\endcsname{3.6\times10^{-9}}
\expandafter\def\csname GQC@hfdft.ref.cpu.force\endcsname{2.2\times10^{-8}}
\expandafter\def\csname GQC@hfdft.ref.grid.energy\endcsname{1.1\times10^{-6}}
\expandafter\def\csname GQC@hfdft.ref.grid.force\endcsname{1.8\times10^{-5}}
\expandafter\def\csname GQC@hfdft.rtx5090.both.wins\endcsname{64}
\expandafter\def\csname GQC@hfdft.rtx5090.cold.median.3\endcsname{3.29}
\expandafter\def\csname GQC@hfdft.rtx5090.cold.median.96\endcsname{1.21}
\expandafter\def\csname GQC@hfdft.rtx5090.cold.wins\endcsname{64}
\expandafter\def\csname GQC@hfdft.rtx5090.faster\endcsname{in all 128 comparisons}
\expandafter\def\csname GQC@hfdft.rtx5090.faster.sized\endcsname{in every cold and warm comparison}
\expandafter\def\csname GQC@hfdft.rtx5090.large.warm.gqc.max\endcsname{23}
\expandafter\def\csname GQC@hfdft.rtx5090.large.warm.gqc.min\endcsname{2.7}
\expandafter\def\csname GQC@hfdft.rtx5090.near\endcsname{the comparisons closest to parity are two 96-atom cold calls (DF HF and DF r$^2$SCAN-3c), with GPU4PySCF-to-GenerativeQC time ratios of 1.003--1.011}
\expandafter\def\csname GQC@hfdft.rtx5090.saved.warm.3\endcsname{0.40}
\expandafter\def\csname GQC@hfdft.rtx5090.saved.warm.96\endcsname{3.4}
\expandafter\def\csname GQC@hfdft.rtx5090.small.warm.gpu4pyscf.ms\endcsname{450}
\expandafter\def\csname GQC@hfdft.rtx5090.small.warm.gqc.ms\endcsname{67}
\expandafter\def\csname GQC@hfdft.rtx5090.total.cold.gpu4pyscf\endcsname{1494}
\expandafter\def\csname GQC@hfdft.rtx5090.total.cold.gqc\endcsname{824}
\expandafter\def\csname GQC@hfdft.rtx5090.total.warm.gpu4pyscf\endcsname{489}
\expandafter\def\csname GQC@hfdft.rtx5090.total.warm.gqc\endcsname{145}
\expandafter\def\csname GQC@hfdft.rtx5090.warm.median.3\endcsname{7.47}
\expandafter\def\csname GQC@hfdft.rtx5090.warm.median.96\endcsname{1.58}
\expandafter\def\csname GQC@hfdft.rtx5090.warm.wins\endcsname{64}
\expandafter\def\csname GQC@hfdft.rtx5090.wbmv.max\endcsname{9.4}
\expandafter\def\csname GQC@hfdft.rtx5090.wbmv.min\endcsname{5.5}
\expandafter\def\csname GQC@hfdft.rtx5090.wbmv96.gpu4pyscf\endcsname{119}
\expandafter\def\csname GQC@hfdft.rtx5090.wbmv96.gqc\endcsname{12.7}
\expandafter\def\csname GQC@hfdft.rtx5090.wbmv96.path\endcsname{Direct}
\expandafter\def\csname GQC@hfdft.small.warm.gpu4pyscf.max\endcsname{2.1}
\expandafter\def\csname GQC@hfdft.small.warm.gpu4pyscf.min\endcsname{0.21}
\expandafter\def\csname GQC@hfdft.small.warm.gqc.max\endcsname{0.29}
\expandafter\def\csname GQC@hfdft.small.warm.gqc.min\endcsname{0.011}
\expandafter\def\csname GQC@rows.cbs.alkanes\endcsname{%
    C1 & 0.0520 & 0.0522 & -- & -- & 0.0543 & 0.0543 & 0.419 & 0.419 \\
    C2 & 0.629 & 0.628 & 10.6 & 10.6 & 0.633 & 0.633 & 0.238 & 0.238 \\
    C3 & 0.648 & 0.647 & -- & -- & 0.651 & 0.651 & 0.256 & 0.256 \\
    C4 & 0.800 & 0.800 & 23.1 & 23.1 & 0.796 & 0.796 & 0.305 & 0.304 \\
    C5 & 1.25 & 1.25 & -- & -- & 1.25 & 1.25 & 0.666 & 0.664 \\
    C6 & 1.75 & 1.75 & -- & -- & 1.74 & 1.74 & 0.796 & 0.798 \\
    C7 & 2.37 & 2.37 & 9.91 & 9.91 & 2.37 & 2.37 & 1.25 & 1.25 \\
}
\expandafter\def\csname GQC@rows.cbs.times\endcsname{%
    RTX~5090 & C1 & 5 & 3.211 & 0.101 & 0.0504 & 0.247 & 0.417 & 1.99 & 0.408 \\
    RTX~5090 & C2 & 8 & 8.597 & 0.462 & 0.404 & 1.47 & 0.776 & 4.58 & 0.901 \\
    RTX~5090 & C3 & 11 & 24.58 & 2.25 & 3.00 & 7.06 & 1.43 & 9.24 & 1.60 \\
    RTX~5090 & C4 & 14 & 67.26 & 8.15 & 12.3 & 24.8 & 2.47 & 16.5 & 3.07 \\
    RTX~5090 & C5 & 17 & 172.4 & 23.9 & 46.7 & 67.1 & 3.92 & 27.5 & 3.23 \\
    RTX~5090 & C6 & 20 & 464.9 & 66.1 & 158 & 185 & 6.31 & 44.4 & 5.07 \\
    RTX~5090 & C7 & 23 & 1019 & 145 & 343 & 449 & 9.53 & 65.4 & 6.38 \\
    \midrule
    H100 & C1 & 5 & 1.138 & 0.0730 & 0.0341 & 0.131 & 0.197 & 0.614 & 0.0887 \\
    H100 & C2 & 8 & 3.202 & 0.315 & 0.201 & 0.598 & 0.340 & 1.56 & 0.189 \\
    H100 & C3 & 11 & 8.892 & 1.23 & 0.813 & 2.54 & 0.593 & 3.42 & 0.302 \\
    H100 & C4 & 14 & 23.04 & 3.82 & 2.93 & 8.28 & 1.04 & 6.50 & 0.471 \\
    H100 & C5 & 17 & 55.34 & 10.4 & 8.53 & 22.3 & 1.79 & 11.6 & 0.698 \\
    H100 & C6 & 20 & 125.0 & 24.7 & 23.7 & 53.4 & 2.86 & 19.2 & 1.06 \\
    H100 & C7 & 23 & 267.3 & 52.9 & 56.0 & 124 & 4.53 & 27.9 & 1.54 \\
}
\expandafter\def\csname GQC@rows.cbs.validation\endcsname{%
    Full-DF & NH$_3$ & 6.03 & 6.03 & 11.5 & 11.5 & 6.40 & 6.40 \\
    Full-DF & NH$_4^+$ & 0.101 & 0.101 & 17.3 & 17.3 & 9.18 & 9.18 \\
    Full-DF & H$_2$CO & 0.614 & 0.613 & 37.2 & 37.2 & 14.8 & 14.8 \\
    Full-DF & H$_2$, stretched & 0.469 & 0.115 & 48.5 & 48.5 & 33.7 & 33.7 \\
    Full-DF & OH$^-$ & 1.20 & 1.20 & 3.80 & 3.80 & 1.87 & 1.87 \\
    Full-DF & OH$^-$, aug-cc-pV$X$Z & 2.84 & 2.85 & 19.0 & 19.0 & 8.93 & 8.93 \\
    Full-DF & N$_2$ & 1.65 & 1.65 & 1.40 & 1.39 & 0.806 & 0.805 \\
    Full-DF & N$_2$, stretched & 20.9 & 5.58 & 113 & 11.9 & 59.1 & 6.73 \\
    Full-DF & H$_2$O & 7.08 & 7.08 & 10.0 & 10.0 & 5.02 & 5.02 \\
    Full-DF & (H$_2$O)$_2$ & 1.14 & 1.14 & 25.1 & 25.1 & 9.00 & 9.00 \\
    Full-DF & H$_2$O, stretched & 14.6 & 14.6 & 22.7 & 22.7 & 10.1 & 10.1 \\
    Correlation-only DF & NH$_4^+$ & 0.100 & 0.100 & 33.5 & 33.5 & 16.6 & 16.6 \\
    Correlation-only DF & H$_2$O, stretched & 14.6 & 14.5 & 14.6 & 14.5 & 8.24 & 8.23 \\
}
\expandafter\def\csname GQC@rows.hfdft.errors\endcsname{%
    Direct & HF & 3 & $<$0.001 & $<$0.001 & 0.125 & 0.599 & 0.0681 & 0.391 \\
    Direct & HF & 12 & $<$0.001 & $<$0.001 & 0.638 & 0.974 & 0.267 & 0.432 \\
    Direct & HF & 48 & $<$0.001 & $<$0.001 & 0.180 & 0.140 & 0.0899 & 0.0577 \\
    Direct & HF & 96 & $<$0.001 & $<$0.001 & 0.301 & 0.152 & 0.130 & 0.0682 \\
    Direct & PBE & 3 & 0.122 & 0.122 & 6.02 & 6.11 & 2.97 & 2.98 \\
    Direct & PBE & 12 & 0.245 & 0.245 & 10.0 & 9.80 & 3.56 & 3.54 \\
    Direct & PBE & 48 & 0.515 & 0.515 & 33.6 & 33.7 & 15.6 & 15.6 \\
    Direct & PBE & 96 & 0.516 & 0.516 & 33.8 & 33.9 & 15.6 & 15.6 \\
    Direct & PBE0 & 3 & 0.0941 & 0.0941 & 4.68 & 4.69 & 2.31 & 2.31 \\
    Direct & PBE0 & 12 & 0.166 & 0.166 & 7.63 & 7.79 & 2.72 & 2.77 \\
    Direct & PBE0 & 48 & 0.297 & 0.297 & 22.5 & 22.6 & 10.9 & 10.9 \\
    Direct & PBE0 & 96 & 0.297 & 0.297 & 22.6 & 22.7 & 10.9 & 10.9 \\
    Direct & B3LYP & 3 & 0.112 & 0.112 & 5.73 & 5.93 & 2.80 & 2.85 \\
    Direct & B3LYP & 12 & 0.251 & 0.251 & 9.35 & 9.47 & 3.44 & 3.56 \\
    Direct & B3LYP & 48 & 0.635 & 0.635 & 25.8 & 26.3 & 12.2 & 12.2 \\
    Direct & B3LYP & 96 & 0.638 & 0.638 & 25.9 & 26.1 & 12.2 & 12.2 \\
    Direct & r$^2$SCAN & 3 & 2.07 & 2.07 & 11.6 & 12.0 & 7.68 & 8.03 \\
    Direct & r$^2$SCAN & 12 & 1.82 & 1.82 & 19.0 & 19.4 & 8.45 & 8.53 \\
    Direct & r$^2$SCAN & 48 & 1.60 & 1.60 & 24.1 & 24.0 & 12.9 & 12.9 \\
    Direct & r$^2$SCAN & 96 & 1.60 & 1.60 & 23.9 & 24.0 & 12.9 & 12.9 \\
    Direct & r$^2$SCAN-3c & 3 & 1.73 & 1.73 & 13.4 & 15.7 & 7.64 & 8.43 \\
    Direct & r$^2$SCAN-3c & 12 & 1.61 & 1.61 & 20.9 & 21.1 & 7.58 & 7.61 \\
    Direct & r$^2$SCAN-3c & 48 & 3.04 & 3.04 & 54.8 & 54.6 & 24.6 & 24.6 \\
    Direct & r$^2$SCAN-3c & 96 & 3.04 & 3.04 & 54.5 & 54.8 & 24.6 & 24.7 \\
    Direct & M06-2X & 3 & 0.304 & 0.304 & 47.6 & 48.1 & 32.8 & 33.2 \\
    Direct & M06-2X & 12 & 0.753 & 0.753 & 52.5 & 52.9 & 26.1 & 26.2 \\
    Direct & M06-2X & 48 & 0.120 & 0.120 & 51.2 & 51.6 & 25.2 & 25.3 \\
    Direct & M06-2X & 96 & 0.119 & 0.119 & 51.2 & 51.4 & 25.2 & 25.3 \\
    Direct & $\omega$B97M-V & 3 & 0.101 & 0.101 & 2.54 & 2.22 & 1.54 & 1.22 \\
    Direct & $\omega$B97M-V & 12 & 0.238 & 0.238 & 7.59 & 6.13 & 2.59 & 1.89 \\
    Direct & $\omega$B97M-V & 48 & 0.486 & 0.486 & 22.0 & 22.9 & 10.6 & 10.5 \\
    Direct & $\omega$B97M-V & 96 & 0.486 & 0.486 & 22.1 & 22.2 & 10.6 & 10.5 \\
    DF & HF & 3 & 18.5 & 18.5 & 9.17 & 8.88 & 4.79 & 4.64 \\
    DF & HF & 12 & 20.3 & 20.3 & 16.6 & 16.5 & 6.31 & 6.31 \\
    DF & HF & 48 & 21.1 & 21.1 & 19.7 & 19.7 & 8.20 & 8.18 \\
    DF & HF & 96 & 21.0 & 21.0 & 19.7 & 19.7 & 8.17 & 8.17 \\
    DF & PBE & 3 & 8.78 & 8.78 & 8.99 & 9.08 & 4.22 & 4.26 \\
    DF & PBE & 12 & 8.01 & 8.01 & 9.80 & 9.49 & 4.28 & 4.22 \\
    DF & PBE & 48 & 7.32 & 7.32 & 36.8 & 36.9 & 15.8 & 15.8 \\
    DF & PBE & 96 & 7.31 & 7.31 & 37.0 & 37.1 & 15.8 & 15.8 \\
    DF & PBE0 & 3 & 1.84 & 1.84 & 7.20 & 7.21 & 3.60 & 3.59 \\
    DF & PBE0 & 12 & 0.778 & 0.778 & 10.5 & 10.4 & 4.12 & 4.15 \\
    DF & PBE0 & 48 & 0.0897 & 0.0897 & 23.0 & 23.2 & 10.1 & 10.1 \\
    DF & PBE0 & 96 & 0.0908 & 0.0908 & 23.1 & 23.3 & 10.1 & 10.1 \\
    DF & B3LYP & 3 & 3.27 & 3.27 & 8.39 & 8.59 & 4.06 & 4.18 \\
    DF & B3LYP & 12 & 2.25 & 2.25 & 11.3 & 11.5 & 4.66 & 4.74 \\
    DF & B3LYP & 48 & 1.36 & 1.36 & 27.1 & 27.6 & 11.5 & 11.5 \\
    DF & B3LYP & 96 & 1.35 & 1.35 & 27.2 & 27.4 & 11.5 & 11.5 \\
    DF & r$^2$SCAN & 3 & 6.83 & 6.83 & 13.2 & 14.3 & 8.49 & 8.91 \\
    DF & r$^2$SCAN & 12 & 6.39 & 6.39 & 18.6 & 19.0 & 8.65 & 8.74 \\
    DF & r$^2$SCAN & 48 & 6.17 & 6.17 & 27.3 & 27.1 & 13.3 & 13.3 \\
    DF & r$^2$SCAN & 96 & 6.17 & 6.17 & 27.2 & 27.2 & 13.4 & 13.4 \\
    DF & r$^2$SCAN-3c & 3 & 0.653 & 0.653 & 13.9 & 16.2 & 8.30 & 9.04 \\
    DF & r$^2$SCAN-3c & 12 & 0.575 & 0.575 & 23.1 & 22.7 & 8.57 & 8.51 \\
    DF & r$^2$SCAN-3c & 48 & 2.04 & 2.04 & 56.3 & 56.2 & 24.7 & 24.7 \\
    DF & r$^2$SCAN-3c & 96 & 2.04 & 2.04 & 56.0 & 56.3 & 24.7 & 24.8 \\
    DF & M06-2X & 3 & 6.18 & 6.18 & 49.7 & 50.3 & 33.1 & 33.5 \\
    DF & M06-2X & 12 & 6.98 & 6.98 & 56.8 & 57.2 & 26.7 & 26.8 \\
    DF & M06-2X & 48 & 8.20 & 8.20 & 56.7 & 56.9 & 25.8 & 25.8 \\
    DF & M06-2X & 96 & 8.20 & 8.20 & 56.7 & 56.9 & 25.8 & 25.8 \\
    DF & $\omega$B97M-V & 3 & 4.68 & 4.68 & 4.62 & 4.73 & 2.06 & 2.07 \\
    DF & $\omega$B97M-V & 12 & 3.74 & 3.74 & 7.21 & 7.22 & 3.12 & 3.06 \\
    DF & $\omega$B97M-V & 48 & 3.01 & 3.01 & 20.0 & 20.7 & 9.41 & 9.38 \\
    DF & $\omega$B97M-V & 96 & 3.01 & 3.01 & 20.0 & 20.0 & 9.43 & 9.40 \\
}
\expandafter\def\csname GQC@rows.hfdft.timing.h100\endcsname{%
    Direct & HF & 3 & \begin{tabular}[t]{@{}l@{}}0.7502\\{\footnotesize[0.7488--0.7563]}\end{tabular} & \begin{tabular}[t]{@{}l@{}}0.01675\\{\footnotesize[0.01662--0.01693]}\end{tabular} & \begin{tabular}[t]{@{}l@{}}5.359\\{\footnotesize[5.325--5.484]}\end{tabular} & \begin{tabular}[t]{@{}l@{}}1.455\\{\footnotesize[1.454--1.495]}\end{tabular} \\
    Direct & HF & 12 & \begin{tabular}[t]{@{}l@{}}0.9996\\{\footnotesize[0.9986--1.015]}\end{tabular} & \begin{tabular}[t]{@{}l@{}}0.02886\\{\footnotesize[0.02835--0.03753]}\end{tabular} & \begin{tabular}[t]{@{}l@{}}5.989\\{\footnotesize[5.907--6.194]}\end{tabular} & \begin{tabular}[t]{@{}l@{}}1.511\\{\footnotesize[1.492--1.543]}\end{tabular} \\
    Direct & HF & 48 & \begin{tabular}[t]{@{}l@{}}4.186\\{\footnotesize[4.165--4.259]}\end{tabular} & \begin{tabular}[t]{@{}l@{}}0.3645\\{\footnotesize[0.3624--0.3646]}\end{tabular} & \begin{tabular}[t]{@{}l@{}}6.859\\{\footnotesize[6.811--6.924]}\end{tabular} & \begin{tabular}[t]{@{}l@{}}1.59\\{\footnotesize[1.588--1.6]}\end{tabular} \\
    Direct & HF & 96 & \begin{tabular}[t]{@{}l@{}}6.19\\{\footnotesize[6.188--6.199]}\end{tabular} & \begin{tabular}[t]{@{}l@{}}1.291\\{\footnotesize[1.29--1.298]}\end{tabular} & \begin{tabular}[t]{@{}l@{}}7.234\\{\footnotesize[7.121--7.723]}\end{tabular} & \begin{tabular}[t]{@{}l@{}}1.665\\{\footnotesize[1.634--1.724]}\end{tabular} \\
    Direct & PBE & 3 & \begin{tabular}[t]{@{}l@{}}1.008\\{\footnotesize[0.9975--1.015]}\end{tabular} & \begin{tabular}[t]{@{}l@{}}0.02112\\{\footnotesize[0.02056--0.02125]}\end{tabular} & \begin{tabular}[t]{@{}l@{}}4.98\\{\footnotesize[4.871--5.006]}\end{tabular} & \begin{tabular}[t]{@{}l@{}}0.6138\\{\footnotesize[0.6131--0.6259]}\end{tabular} \\
    Direct & PBE & 12 & \begin{tabular}[t]{@{}l@{}}1.199\\{\footnotesize[1.189--1.215]}\end{tabular} & \begin{tabular}[t]{@{}l@{}}0.09924\\{\footnotesize[0.09798--0.09946]}\end{tabular} & \begin{tabular}[t]{@{}l@{}}5.437\\{\footnotesize[5.343--5.492]}\end{tabular} & \begin{tabular}[t]{@{}l@{}}0.73\\{\footnotesize[0.7257--0.7307]}\end{tabular} \\
    Direct & PBE & 48 & \begin{tabular}[t]{@{}l@{}}2.614\\{\footnotesize[2.598--2.628]}\end{tabular} & \begin{tabular}[t]{@{}l@{}}0.4268\\{\footnotesize[0.4254--0.4271]}\end{tabular} & \begin{tabular}[t]{@{}l@{}}6.715\\{\footnotesize[6.672--6.723]}\end{tabular} & \begin{tabular}[t]{@{}l@{}}1.18\\{\footnotesize[1.17--1.181]}\end{tabular} \\
    Direct & PBE & 96 & \begin{tabular}[t]{@{}l@{}}4.203\\{\footnotesize[4.201--4.229]}\end{tabular} & \begin{tabular}[t]{@{}l@{}}1.069\\{\footnotesize[1.069--1.072]}\end{tabular} & \begin{tabular}[t]{@{}l@{}}8.434\\{\footnotesize[8.405--8.561]}\end{tabular} & \begin{tabular}[t]{@{}l@{}}1.98\\{\footnotesize[1.898--1.989]}\end{tabular} \\
    Direct & PBE0 & 3 & \begin{tabular}[t]{@{}l@{}}1.048\\{\footnotesize[1.037--1.079]}\end{tabular} & \begin{tabular}[t]{@{}l@{}}0.02793\\{\footnotesize[0.02786--0.02817]}\end{tabular} & \begin{tabular}[t]{@{}l@{}}8.589\\{\footnotesize[8.531--8.67]}\end{tabular} & \begin{tabular}[t]{@{}l@{}}1.529\\{\footnotesize[1.527--1.534]}\end{tabular} \\
    Direct & PBE0 & 12 & \begin{tabular}[t]{@{}l@{}}1.458\\{\footnotesize[1.444--1.472]}\end{tabular} & \begin{tabular}[t]{@{}l@{}}0.1402\\{\footnotesize[0.1394--0.1408]}\end{tabular} & \begin{tabular}[t]{@{}l@{}}9.981\\{\footnotesize[9.971--10.05]}\end{tabular} & \begin{tabular}[t]{@{}l@{}}1.667\\{\footnotesize[1.666--1.703]}\end{tabular} \\
    Direct & PBE0 & 48 & \begin{tabular}[t]{@{}l@{}}3.829\\{\footnotesize[3.826--3.838]}\end{tabular} & \begin{tabular}[t]{@{}l@{}}0.7314\\{\footnotesize[0.7284--0.7321]}\end{tabular} & \begin{tabular}[t]{@{}l@{}}11.09\\{\footnotesize[10.99--11.21]}\end{tabular} & \begin{tabular}[t]{@{}l@{}}2.092\\{\footnotesize[2.085--2.126]}\end{tabular} \\
    Direct & PBE0 & 96 & \begin{tabular}[t]{@{}l@{}}6.762\\{\footnotesize[6.746--6.991]}\end{tabular} & \begin{tabular}[t]{@{}l@{}}1.775\\{\footnotesize[1.767--1.781]}\end{tabular} & \begin{tabular}[t]{@{}l@{}}12.82\\{\footnotesize[12.75--12.84]}\end{tabular} & \begin{tabular}[t]{@{}l@{}}2.847\\{\footnotesize[2.834--2.93]}\end{tabular} \\
    Direct & B3LYP & 3 & \begin{tabular}[t]{@{}l@{}}1.058\\{\footnotesize[1.055--1.078]}\end{tabular} & \begin{tabular}[t]{@{}l@{}}0.02979\\{\footnotesize[0.02944--0.03058]}\end{tabular} & \begin{tabular}[t]{@{}l@{}}8.625\\{\footnotesize[8.594--8.676]}\end{tabular} & \begin{tabular}[t]{@{}l@{}}1.544\\{\footnotesize[1.543--1.579]}\end{tabular} \\
    Direct & B3LYP & 12 & \begin{tabular}[t]{@{}l@{}}1.47\\{\footnotesize[1.459--1.477]}\end{tabular} & \begin{tabular}[t]{@{}l@{}}0.143\\{\footnotesize[0.1424--0.1435]}\end{tabular} & \begin{tabular}[t]{@{}l@{}}10.09\\{\footnotesize[9.888--10.35]}\end{tabular} & \begin{tabular}[t]{@{}l@{}}1.679\\{\footnotesize[1.675--1.68]}\end{tabular} \\
    Direct & B3LYP & 48 & \begin{tabular}[t]{@{}l@{}}3.869\\{\footnotesize[3.862--3.909]}\end{tabular} & \begin{tabular}[t]{@{}l@{}}0.7433\\{\footnotesize[0.7403--0.7444]}\end{tabular} & \begin{tabular}[t]{@{}l@{}}11.34\\{\footnotesize[11.08--11.34]}\end{tabular} & \begin{tabular}[t]{@{}l@{}}2.13\\{\footnotesize[2.081--2.137]}\end{tabular} \\
    Direct & B3LYP & 96 & \begin{tabular}[t]{@{}l@{}}6.836\\{\footnotesize[6.831--6.851]}\end{tabular} & \begin{tabular}[t]{@{}l@{}}1.786\\{\footnotesize[1.782--1.795]}\end{tabular} & \begin{tabular}[t]{@{}l@{}}13.29\\{\footnotesize[13.19--13.75]}\end{tabular} & \begin{tabular}[t]{@{}l@{}}2.87\\{\footnotesize[2.804--2.918]}\end{tabular} \\
    Direct & r$^2$SCAN & 3 & \begin{tabular}[t]{@{}l@{}}1.108\\{\footnotesize[1.103--1.11]}\end{tabular} & \begin{tabular}[t]{@{}l@{}}0.03466\\{\footnotesize[0.03365--0.0357]}\end{tabular} & \begin{tabular}[t]{@{}l@{}}4.915\\{\footnotesize[4.894--5.15]}\end{tabular} & \begin{tabular}[t]{@{}l@{}}0.626\\{\footnotesize[0.6244--0.6703]}\end{tabular} \\
    Direct & r$^2$SCAN & 12 & \begin{tabular}[t]{@{}l@{}}1.304\\{\footnotesize[1.295--1.321]}\end{tabular} & \begin{tabular}[t]{@{}l@{}}0.1248\\{\footnotesize[0.1241--0.1249]}\end{tabular} & \begin{tabular}[t]{@{}l@{}}5.415\\{\footnotesize[5.371--5.524]}\end{tabular} & \begin{tabular}[t]{@{}l@{}}0.8023\\{\footnotesize[0.7926--0.8357]}\end{tabular} \\
    Direct & r$^2$SCAN & 48 & \begin{tabular}[t]{@{}l@{}}3.084\\{\footnotesize[3.059--3.086]}\end{tabular} & \begin{tabular}[t]{@{}l@{}}0.5312\\{\footnotesize[0.5312--0.534]}\end{tabular} & \begin{tabular}[t]{@{}l@{}}7.321\\{\footnotesize[7.309--7.479]}\end{tabular} & \begin{tabular}[t]{@{}l@{}}1.447\\{\footnotesize[1.439--1.464]}\end{tabular} \\
    Direct & r$^2$SCAN & 96 & \begin{tabular}[t]{@{}l@{}}4.908\\{\footnotesize[4.907--4.915]}\end{tabular} & \begin{tabular}[t]{@{}l@{}}1.186\\{\footnotesize[1.174--1.189]}\end{tabular} & \begin{tabular}[t]{@{}l@{}}9.769\\{\footnotesize[9.71--9.791]}\end{tabular} & \begin{tabular}[t]{@{}l@{}}2.479\\{\footnotesize[2.455--2.482]}\end{tabular} \\
    Direct & r$^2$SCAN-3c & 3 & \begin{tabular}[t]{@{}l@{}}1.122\\{\footnotesize[1.101--1.126]}\end{tabular} & \begin{tabular}[t]{@{}l@{}}0.05059\\{\footnotesize[0.05035--0.0508]}\end{tabular} & \begin{tabular}[t]{@{}l@{}}5.606\\{\footnotesize[5.583--5.638]}\end{tabular} & \begin{tabular}[t]{@{}l@{}}1.015\\{\footnotesize[1.006--1.063]}\end{tabular} \\
    Direct & r$^2$SCAN-3c & 12 & \begin{tabular}[t]{@{}l@{}}1.857\\{\footnotesize[1.808--1.862]}\end{tabular} & \begin{tabular}[t]{@{}l@{}}0.456\\{\footnotesize[0.4554--0.4629]}\end{tabular} & \begin{tabular}[t]{@{}l@{}}6.296\\{\footnotesize[6.216--6.418]}\end{tabular} & \begin{tabular}[t]{@{}l@{}}1.208\\{\footnotesize[1.202--1.25]}\end{tabular} \\
    Direct & r$^2$SCAN-3c & 48 & \begin{tabular}[t]{@{}l@{}}3.631\\{\footnotesize[3.603--3.632]}\end{tabular} & \begin{tabular}[t]{@{}l@{}}1.201\\{\footnotesize[1.201--1.216]}\end{tabular} & \begin{tabular}[t]{@{}l@{}}9.278\\{\footnotesize[9.071--9.412]}\end{tabular} & \begin{tabular}[t]{@{}l@{}}2.116\\{\footnotesize[2.108--2.119]}\end{tabular} \\
    Direct & r$^2$SCAN-3c & 96 & \begin{tabular}[t]{@{}l@{}}9.342\\{\footnotesize[9.339--9.345]}\end{tabular} & \begin{tabular}[t]{@{}l@{}}3.485\\{\footnotesize[3.484--3.496]}\end{tabular} & \begin{tabular}[t]{@{}l@{}}12.32\\{\footnotesize[12.23--12.35]}\end{tabular} & \begin{tabular}[t]{@{}l@{}}3.788\\{\footnotesize[3.771--3.91]}\end{tabular} \\
    Direct & M06-2X & 3 & \begin{tabular}[t]{@{}l@{}}1.101\\{\footnotesize[1.097--1.109]}\end{tabular} & \begin{tabular}[t]{@{}l@{}}0.03604\\{\footnotesize[0.03541--0.03708]}\end{tabular} & \begin{tabular}[t]{@{}l@{}}8.718\\{\footnotesize[8.609--8.769]}\end{tabular} & \begin{tabular}[t]{@{}l@{}}1.557\\{\footnotesize[1.545--1.584]}\end{tabular} \\
    Direct & M06-2X & 12 & \begin{tabular}[t]{@{}l@{}}1.495\\{\footnotesize[1.493--1.506]}\end{tabular} & \begin{tabular}[t]{@{}l@{}}0.1519\\{\footnotesize[0.1516--0.1523]}\end{tabular} & \begin{tabular}[t]{@{}l@{}}10.73\\{\footnotesize[10.47--10.89]}\end{tabular} & \begin{tabular}[t]{@{}l@{}}1.732\\{\footnotesize[1.693--1.74]}\end{tabular} \\
    Direct & M06-2X & 48 & \begin{tabular}[t]{@{}l@{}}4.736\\{\footnotesize[4.732--4.768]}\end{tabular} & \begin{tabular}[t]{@{}l@{}}0.8205\\{\footnotesize[0.8165--0.8244]}\end{tabular} & \begin{tabular}[t]{@{}l@{}}11.87\\{\footnotesize[11.86--11.93]}\end{tabular} & \begin{tabular}[t]{@{}l@{}}2.382\\{\footnotesize[2.378--2.386]}\end{tabular} \\
    Direct & M06-2X & 96 & \begin{tabular}[t]{@{}l@{}}8.018\\{\footnotesize[7.967--8.033]}\end{tabular} & \begin{tabular}[t]{@{}l@{}}1.836\\{\footnotesize[1.825--1.838]}\end{tabular} & \begin{tabular}[t]{@{}l@{}}15.03\\{\footnotesize[14.9--15.44]}\end{tabular} & \begin{tabular}[t]{@{}l@{}}3.416\\{\footnotesize[3.387--3.466]}\end{tabular} \\
    Direct & $\omega$B97M-V & 3 & \begin{tabular}[t]{@{}l@{}}1.579\\{\footnotesize[1.572--1.585]}\end{tabular} & \begin{tabular}[t]{@{}l@{}}0.1108\\{\footnotesize[0.1104--0.1108]}\end{tabular} & \begin{tabular}[t]{@{}l@{}}10.07\\{\footnotesize[9.82--10.18]}\end{tabular} & \begin{tabular}[t]{@{}l@{}}2.107\\{\footnotesize[2.081--2.142]}\end{tabular} \\
    Direct & $\omega$B97M-V & 12 & \begin{tabular}[t]{@{}l@{}}2.296\\{\footnotesize[2.292--2.317]}\end{tabular} & \begin{tabular}[t]{@{}l@{}}0.359\\{\footnotesize[0.3581--0.3631]}\end{tabular} & \begin{tabular}[t]{@{}l@{}}12.18\\{\footnotesize[12.06--13.05]}\end{tabular} & \begin{tabular}[t]{@{}l@{}}2.484\\{\footnotesize[2.439--2.59]}\end{tabular} \\
    Direct & $\omega$B97M-V & 48 & \begin{tabular}[t]{@{}l@{}}13.19\\{\footnotesize[13.13--13.25]}\end{tabular} & \begin{tabular}[t]{@{}l@{}}2.458\\{\footnotesize[2.45--2.461]}\end{tabular} & \begin{tabular}[t]{@{}l@{}}19.49\\{\footnotesize[19.34--19.57]}\end{tabular} & \begin{tabular}[t]{@{}l@{}}5.286\\{\footnotesize[5.275--5.328]}\end{tabular} \\
    Direct & $\omega$B97M-V & 96 & \begin{tabular}[t]{@{}l@{}}33.47\\{\footnotesize[33.33--33.59]}\end{tabular} & \begin{tabular}[t]{@{}l@{}}4.268\\{\footnotesize[4.258--4.269]}\end{tabular} & \begin{tabular}[t]{@{}l@{}}41.56\\{\footnotesize[41.49--41.58]}\end{tabular} & \begin{tabular}[t]{@{}l@{}}12.61\\{\footnotesize[12.51--12.63]}\end{tabular} \\
    DF & HF & 3 & \begin{tabular}[t]{@{}l@{}}0.5031\\{\footnotesize[0.4955--0.5075]}\end{tabular} & \begin{tabular}[t]{@{}l@{}}0.01115\\{\footnotesize[0.01109--0.01142]}\end{tabular} & \begin{tabular}[t]{@{}l@{}}4.274\\{\footnotesize[4.22--4.302]}\end{tabular} & \begin{tabular}[t]{@{}l@{}}0.2068\\{\footnotesize[0.2066--0.2101]}\end{tabular} \\
    DF & HF & 12 & \begin{tabular}[t]{@{}l@{}}0.6705\\{\footnotesize[0.6676--0.6733]}\end{tabular} & \begin{tabular}[t]{@{}l@{}}0.02743\\{\footnotesize[0.02737--0.02768]}\end{tabular} & \begin{tabular}[t]{@{}l@{}}4.316\\{\footnotesize[4.292--4.54]}\end{tabular} & \begin{tabular}[t]{@{}l@{}}0.2227\\{\footnotesize[0.2212--0.2262]}\end{tabular} \\
    DF & HF & 48 & \begin{tabular}[t]{@{}l@{}}1.092\\{\footnotesize[1.074--1.182]}\end{tabular} & \begin{tabular}[t]{@{}l@{}}0.1299\\{\footnotesize[0.1292--0.1302]}\end{tabular} & \begin{tabular}[t]{@{}l@{}}4.479\\{\footnotesize[4.439--4.721]}\end{tabular} & \begin{tabular}[t]{@{}l@{}}0.336\\{\footnotesize[0.3324--0.3382]}\end{tabular} \\
    DF & HF & 96 & \begin{tabular}[t]{@{}l@{}}3.012\\{\footnotesize[3.003--3.032]}\end{tabular} & \begin{tabular}[t]{@{}l@{}}0.6418\\{\footnotesize[0.6408--0.6504]}\end{tabular} & \begin{tabular}[t]{@{}l@{}}5.321\\{\footnotesize[5.31--5.384]}\end{tabular} & \begin{tabular}[t]{@{}l@{}}0.7652\\{\footnotesize[0.7647--0.7655]}\end{tabular} \\
    DF & PBE & 3 & \begin{tabular}[t]{@{}l@{}}0.693\\{\footnotesize[0.6926--0.7073]}\end{tabular} & \begin{tabular}[t]{@{}l@{}}0.0238\\{\footnotesize[0.02363--0.02385]}\end{tabular} & \begin{tabular}[t]{@{}l@{}}4.633\\{\footnotesize[4.604--4.815]}\end{tabular} & \begin{tabular}[t]{@{}l@{}}0.214\\{\footnotesize[0.214--0.2147]}\end{tabular} \\
    DF & PBE & 12 & \begin{tabular}[t]{@{}l@{}}0.9616\\{\footnotesize[0.9567--0.9633]}\end{tabular} & \begin{tabular}[t]{@{}l@{}}0.06278\\{\footnotesize[0.06271--0.06307]}\end{tabular} & \begin{tabular}[t]{@{}l@{}}5.097\\{\footnotesize[4.865--5.151]}\end{tabular} & \begin{tabular}[t]{@{}l@{}}0.3107\\{\footnotesize[0.3055--0.3109]}\end{tabular} \\
    DF & PBE & 48 & \begin{tabular}[t]{@{}l@{}}2.315\\{\footnotesize[2.313--2.316]}\end{tabular} & \begin{tabular}[t]{@{}l@{}}0.3292\\{\footnotesize[0.3269--0.3305]}\end{tabular} & \begin{tabular}[t]{@{}l@{}}5.973\\{\footnotesize[5.913--6.225]}\end{tabular} & \begin{tabular}[t]{@{}l@{}}0.7602\\{\footnotesize[0.7559--0.7725]}\end{tabular} \\
    DF & PBE & 96 & \begin{tabular}[t]{@{}l@{}}3.911\\{\footnotesize[3.898--3.912]}\end{tabular} & \begin{tabular}[t]{@{}l@{}}0.8803\\{\footnotesize[0.8758--0.8817]}\end{tabular} & \begin{tabular}[t]{@{}l@{}}7.884\\{\footnotesize[7.627--7.974]}\end{tabular} & \begin{tabular}[t]{@{}l@{}}1.594\\{\footnotesize[1.55--1.614]}\end{tabular} \\
    DF & PBE0 & 3 & \begin{tabular}[t]{@{}l@{}}0.7047\\{\footnotesize[0.6972--0.706]}\end{tabular} & \begin{tabular}[t]{@{}l@{}}0.02738\\{\footnotesize[0.02718--0.02765]}\end{tabular} & \begin{tabular}[t]{@{}l@{}}4.533\\{\footnotesize[4.473--4.555]}\end{tabular} & \begin{tabular}[t]{@{}l@{}}0.2644\\{\footnotesize[0.2632--0.2668]}\end{tabular} \\
    DF & PBE0 & 12 & \begin{tabular}[t]{@{}l@{}}0.9735\\{\footnotesize[0.9667--0.9977]}\end{tabular} & \begin{tabular}[t]{@{}l@{}}0.07201\\{\footnotesize[0.07185--0.07253]}\end{tabular} & \begin{tabular}[t]{@{}l@{}}4.715\\{\footnotesize[4.708--4.716]}\end{tabular} & \begin{tabular}[t]{@{}l@{}}0.3646\\{\footnotesize[0.3567--0.3686]}\end{tabular} \\
    DF & PBE0 & 48 & \begin{tabular}[t]{@{}l@{}}2.344\\{\footnotesize[2.337--2.368]}\end{tabular} & \begin{tabular}[t]{@{}l@{}}0.3498\\{\footnotesize[0.3497--0.3498]}\end{tabular} & \begin{tabular}[t]{@{}l@{}}5.751\\{\footnotesize[5.713--5.77]}\end{tabular} & \begin{tabular}[t]{@{}l@{}}0.8571\\{\footnotesize[0.8451--0.8647]}\end{tabular} \\
    DF & PBE0 & 96 & \begin{tabular}[t]{@{}l@{}}4.772\\{\footnotesize[4.757--4.791]}\end{tabular} & \begin{tabular}[t]{@{}l@{}}1.116\\{\footnotesize[1.113--1.122]}\end{tabular} & \begin{tabular}[t]{@{}l@{}}7.835\\{\footnotesize[7.755--8.109]}\end{tabular} & \begin{tabular}[t]{@{}l@{}}1.945\\{\footnotesize[1.939--1.949]}\end{tabular} \\
    DF & B3LYP & 3 & \begin{tabular}[t]{@{}l@{}}0.7091\\{\footnotesize[0.7038--0.7265]}\end{tabular} & \begin{tabular}[t]{@{}l@{}}0.02844\\{\footnotesize[0.02841--0.02879]}\end{tabular} & \begin{tabular}[t]{@{}l@{}}4.479\\{\footnotesize[4.465--4.522]}\end{tabular} & \begin{tabular}[t]{@{}l@{}}0.2626\\{\footnotesize[0.26--0.2704]}\end{tabular} \\
    DF & B3LYP & 12 & \begin{tabular}[t]{@{}l@{}}0.987\\{\footnotesize[0.9784--0.9877]}\end{tabular} & \begin{tabular}[t]{@{}l@{}}0.07716\\{\footnotesize[0.07667--0.07754]}\end{tabular} & \begin{tabular}[t]{@{}l@{}}4.741\\{\footnotesize[4.718--4.758]}\end{tabular} & \begin{tabular}[t]{@{}l@{}}0.3613\\{\footnotesize[0.3586--0.3634]}\end{tabular} \\
    DF & B3LYP & 48 & \begin{tabular}[t]{@{}l@{}}2.371\\{\footnotesize[2.36--2.4]}\end{tabular} & \begin{tabular}[t]{@{}l@{}}0.358\\{\footnotesize[0.3579--0.3609]}\end{tabular} & \begin{tabular}[t]{@{}l@{}}5.768\\{\footnotesize[5.721--5.819]}\end{tabular} & \begin{tabular}[t]{@{}l@{}}0.8476\\{\footnotesize[0.8457--0.8723]}\end{tabular} \\
    DF & B3LYP & 96 & \begin{tabular}[t]{@{}l@{}}4.798\\{\footnotesize[4.793--4.799]}\end{tabular} & \begin{tabular}[t]{@{}l@{}}1.137\\{\footnotesize[1.134--1.139]}\end{tabular} & \begin{tabular}[t]{@{}l@{}}8.159\\{\footnotesize[7.995--8.21]}\end{tabular} & \begin{tabular}[t]{@{}l@{}}1.97\\{\footnotesize[1.97--1.998]}\end{tabular} \\
    DF & r$^2$SCAN & 3 & \begin{tabular}[t]{@{}l@{}}0.7815\\{\footnotesize[0.7804--0.7819]}\end{tabular} & \begin{tabular}[t]{@{}l@{}}0.03831\\{\footnotesize[0.03795--0.03884]}\end{tabular} & \begin{tabular}[t]{@{}l@{}}4.515\\{\footnotesize[4.509--4.534]}\end{tabular} & \begin{tabular}[t]{@{}l@{}}0.2308\\{\footnotesize[0.2282--0.2322]}\end{tabular} \\
    DF & r$^2$SCAN & 12 & \begin{tabular}[t]{@{}l@{}}1.069\\{\footnotesize[1.063--1.105]}\end{tabular} & \begin{tabular}[t]{@{}l@{}}0.09145\\{\footnotesize[0.09107--0.09192]}\end{tabular} & \begin{tabular}[t]{@{}l@{}}4.984\\{\footnotesize[4.911--4.993]}\end{tabular} & \begin{tabular}[t]{@{}l@{}}0.3735\\{\footnotesize[0.37--0.3792]}\end{tabular} \\
    DF & r$^2$SCAN & 48 & \begin{tabular}[t]{@{}l@{}}2.79\\{\footnotesize[2.779--2.809]}\end{tabular} & \begin{tabular}[t]{@{}l@{}}0.4385\\{\footnotesize[0.4385--0.4396]}\end{tabular} & \begin{tabular}[t]{@{}l@{}}6.613\\{\footnotesize[6.587--6.651]}\end{tabular} & \begin{tabular}[t]{@{}l@{}}1.027\\{\footnotesize[1.022--1.032]}\end{tabular} \\
    DF & r$^2$SCAN & 96 & \begin{tabular}[t]{@{}l@{}}4.624\\{\footnotesize[4.611--4.642]}\end{tabular} & \begin{tabular}[t]{@{}l@{}}0.9897\\{\footnotesize[0.9884--0.9922]}\end{tabular} & \begin{tabular}[t]{@{}l@{}}9.039\\{\footnotesize[9.018--9.07]}\end{tabular} & \begin{tabular}[t]{@{}l@{}}2.109\\{\footnotesize[2.102--2.133]}\end{tabular} \\
    DF & r$^2$SCAN-3c & 3 & \begin{tabular}[t]{@{}l@{}}0.7858\\{\footnotesize[0.7746--0.7953]}\end{tabular} & \begin{tabular}[t]{@{}l@{}}0.0449\\{\footnotesize[0.04432--0.04567]}\end{tabular} & \begin{tabular}[t]{@{}l@{}}4.618\\{\footnotesize[4.586--4.698]}\end{tabular} & \begin{tabular}[t]{@{}l@{}}0.25\\{\footnotesize[0.2451--0.2509]}\end{tabular} \\
    DF & r$^2$SCAN-3c & 12 & \begin{tabular}[t]{@{}l@{}}1.281\\{\footnotesize[1.271--1.288]}\end{tabular} & \begin{tabular}[t]{@{}l@{}}0.1124\\{\footnotesize[0.1122--0.1134]}\end{tabular} & \begin{tabular}[t]{@{}l@{}}5.09\\{\footnotesize[5.084--5.306]}\end{tabular} & \begin{tabular}[t]{@{}l@{}}0.4064\\{\footnotesize[0.4057--0.4091]}\end{tabular} \\
    DF & r$^2$SCAN-3c & 48 & \begin{tabular}[t]{@{}l@{}}2.835\\{\footnotesize[2.827--2.872]}\end{tabular} & \begin{tabular}[t]{@{}l@{}}0.6248\\{\footnotesize[0.6233--0.6261]}\end{tabular} & \begin{tabular}[t]{@{}l@{}}6.979\\{\footnotesize[6.902--7.169]}\end{tabular} & \begin{tabular}[t]{@{}l@{}}1.193\\{\footnotesize[1.184--1.207]}\end{tabular} \\
    DF & r$^2$SCAN-3c & 96 & \begin{tabular}[t]{@{}l@{}}7.498\\{\footnotesize[7.495--7.513]}\end{tabular} & \begin{tabular}[t]{@{}l@{}}1.481\\{\footnotesize[1.472--1.482]}\end{tabular} & \begin{tabular}[t]{@{}l@{}}10.38\\{\footnotesize[10.37--10.46]}\end{tabular} & \begin{tabular}[t]{@{}l@{}}2.648\\{\footnotesize[2.628--2.657]}\end{tabular} \\
    DF & M06-2X & 3 & \begin{tabular}[t]{@{}l@{}}0.7461\\{\footnotesize[0.7412--0.7692]}\end{tabular} & \begin{tabular}[t]{@{}l@{}}0.03557\\{\footnotesize[0.03523--0.03578]}\end{tabular} & \begin{tabular}[t]{@{}l@{}}4.644\\{\footnotesize[4.543--4.696]}\end{tabular} & \begin{tabular}[t]{@{}l@{}}0.2813\\{\footnotesize[0.2808--0.286]}\end{tabular} \\
    DF & M06-2X & 12 & \begin{tabular}[t]{@{}l@{}}1.031\\{\footnotesize[1.026--1.033]}\end{tabular} & \begin{tabular}[t]{@{}l@{}}0.08683\\{\footnotesize[0.08643--0.08782]}\end{tabular} & \begin{tabular}[t]{@{}l@{}}5.044\\{\footnotesize[4.932--5.158]}\end{tabular} & \begin{tabular}[t]{@{}l@{}}0.4309\\{\footnotesize[0.4267--0.4376]}\end{tabular} \\
    DF & M06-2X & 48 & \begin{tabular}[t]{@{}l@{}}3.031\\{\footnotesize[3.021--3.051]}\end{tabular} & \begin{tabular}[t]{@{}l@{}}0.437\\{\footnotesize[0.4368--0.4386]}\end{tabular} & \begin{tabular}[t]{@{}l@{}}6.592\\{\footnotesize[6.523--6.986]}\end{tabular} & \begin{tabular}[t]{@{}l@{}}1.129\\{\footnotesize[1.128--1.149]}\end{tabular} \\
    DF & M06-2X & 96 & \begin{tabular}[t]{@{}l@{}}5.555\\{\footnotesize[5.547--5.581]}\end{tabular} & \begin{tabular}[t]{@{}l@{}}1.189\\{\footnotesize[1.187--1.19]}\end{tabular} & \begin{tabular}[t]{@{}l@{}}9.548\\{\footnotesize[9.495--9.582]}\end{tabular} & \begin{tabular}[t]{@{}l@{}}2.519\\{\footnotesize[2.481--2.527]}\end{tabular} \\
    DF & $\omega$B97M-V & 3 & \begin{tabular}[t]{@{}l@{}}0.9791\\{\footnotesize[0.9754--1.047]}\end{tabular} & \begin{tabular}[t]{@{}l@{}}0.1045\\{\footnotesize[0.104--0.1046]}\end{tabular} & \begin{tabular}[t]{@{}l@{}}5.015\\{\footnotesize[4.907--5.143]}\end{tabular} & \begin{tabular}[t]{@{}l@{}}0.4802\\{\footnotesize[0.4746--0.4839]}\end{tabular} \\
    DF & $\omega$B97M-V & 12 & \begin{tabular}[t]{@{}l@{}}2.588\\{\footnotesize[2.549--2.639]}\end{tabular} & \begin{tabular}[t]{@{}l@{}}0.3029\\{\footnotesize[0.3027--0.3038]}\end{tabular} & \begin{tabular}[t]{@{}l@{}}6.193\\{\footnotesize[6.157--6.211]}\end{tabular} & \begin{tabular}[t]{@{}l@{}}0.872\\{\footnotesize[0.871--0.8785]}\end{tabular} \\
    DF & $\omega$B97M-V & 48 & \begin{tabular}[t]{@{}l@{}}8.725\\{\footnotesize[8.69--8.77]}\end{tabular} & \begin{tabular}[t]{@{}l@{}}1.703\\{\footnotesize[1.696--1.711]}\end{tabular} & \begin{tabular}[t]{@{}l@{}}13.74\\{\footnotesize[13.71--13.9]}\end{tabular} & \begin{tabular}[t]{@{}l@{}}3.819\\{\footnotesize[3.793--3.822]}\end{tabular} \\
    DF & $\omega$B97M-V & 96 & \begin{tabular}[t]{@{}l@{}}20.29\\{\footnotesize[20.21--20.3]}\end{tabular} & \begin{tabular}[t]{@{}l@{}}4.378\\{\footnotesize[4.371--4.397]}\end{tabular} & \begin{tabular}[t]{@{}l@{}}36.76\\{\footnotesize[36.58--37.03]}\end{tabular} & \begin{tabular}[t]{@{}l@{}}11.94\\{\footnotesize[11.89--11.99]}\end{tabular} \\
}
\expandafter\def\csname GQC@rows.hfdft.timing.rtx5090\endcsname{%
    Direct & HF & 3 & \begin{tabular}[t]{@{}l@{}}1.293\\{\footnotesize[1.293--1.295]}\end{tabular} & \begin{tabular}[t]{@{}l@{}}0.07306\\{\footnotesize[0.0724--0.07317]}\end{tabular} & \begin{tabular}[t]{@{}l@{}}3.143\\{\footnotesize[3.129--3.145]}\end{tabular} & \begin{tabular}[t]{@{}l@{}}0.8058\\{\footnotesize[0.8038--0.8076]}\end{tabular} \\
    Direct & HF & 12 & \begin{tabular}[t]{@{}l@{}}1.406\\{\footnotesize[1.406--1.41]}\end{tabular} & \begin{tabular}[t]{@{}l@{}}0.1184\\{\footnotesize[0.1176--0.1197]}\end{tabular} & \begin{tabular}[t]{@{}l@{}}4.109\\{\footnotesize[4.095--4.117]}\end{tabular} & \begin{tabular}[t]{@{}l@{}}0.9926\\{\footnotesize[0.991--0.9949]}\end{tabular} \\
    Direct & HF & 48 & \begin{tabular}[t]{@{}l@{}}6.276\\{\footnotesize[6.273--6.332]}\end{tabular} & \begin{tabular}[t]{@{}l@{}}0.7433\\{\footnotesize[0.7429--0.744]}\end{tabular} & \begin{tabular}[t]{@{}l@{}}7.788\\{\footnotesize[7.781--7.79]}\end{tabular} & \begin{tabular}[t]{@{}l@{}}1.994\\{\footnotesize[1.993--1.995]}\end{tabular} \\
    Direct & HF & 96 & \begin{tabular}[t]{@{}l@{}}11.11\\{\footnotesize[11.11--11.12]}\end{tabular} & \begin{tabular}[t]{@{}l@{}}2.68\\{\footnotesize[2.678--2.684]}\end{tabular} & \begin{tabular}[t]{@{}l@{}}11.67\\{\footnotesize[11.67--11.68]}\end{tabular} & \begin{tabular}[t]{@{}l@{}}3.34\\{\footnotesize[3.339--3.341]}\end{tabular} \\
    Direct & PBE & 3 & \begin{tabular}[t]{@{}l@{}}1.111\\{\footnotesize[1.11--1.112]}\end{tabular} & \begin{tabular}[t]{@{}l@{}}0.03525\\{\footnotesize[0.03523--0.03532]}\end{tabular} & \begin{tabular}[t]{@{}l@{}}3.162\\{\footnotesize[3.159--3.171]}\end{tabular} & \begin{tabular}[t]{@{}l@{}}0.4386\\{\footnotesize[0.4369--0.4402]}\end{tabular} \\
    Direct & PBE & 12 & \begin{tabular}[t]{@{}l@{}}1.459\\{\footnotesize[1.45--1.46]}\end{tabular} & \begin{tabular}[t]{@{}l@{}}0.1861\\{\footnotesize[0.1849--0.1916]}\end{tabular} & \begin{tabular}[t]{@{}l@{}}4.655\\{\footnotesize[4.653--4.661]}\end{tabular} & \begin{tabular}[t]{@{}l@{}}0.7493\\{\footnotesize[0.7492--0.7495]}\end{tabular} \\
    Direct & PBE & 48 & \begin{tabular}[t]{@{}l@{}}4.465\\{\footnotesize[4.464--4.471]}\end{tabular} & \begin{tabular}[t]{@{}l@{}}1.036\\{\footnotesize[1.035--1.037]}\end{tabular} & \begin{tabular}[t]{@{}l@{}}9.043\\{\footnotesize[9.038--9.071]}\end{tabular} & \begin{tabular}[t]{@{}l@{}}2.352\\{\footnotesize[2.349--2.356]}\end{tabular} \\
    Direct & PBE & 96 & \begin{tabular}[t]{@{}l@{}}10.69\\{\footnotesize[10.69--10.71]}\end{tabular} & \begin{tabular}[t]{@{}l@{}}3.735\\{\footnotesize[3.733--3.737]}\end{tabular} & \begin{tabular}[t]{@{}l@{}}15.67\\{\footnotesize[15.62--15.68]}\end{tabular} & \begin{tabular}[t]{@{}l@{}}5.732\\{\footnotesize[5.729--5.733]}\end{tabular} \\
    Direct & PBE0 & 3 & \begin{tabular}[t]{@{}l@{}}1.41\\{\footnotesize[1.409--1.461]}\end{tabular} & \begin{tabular}[t]{@{}l@{}}0.06655\\{\footnotesize[0.06655--0.06671]}\end{tabular} & \begin{tabular}[t]{@{}l@{}}4.927\\{\footnotesize[4.927--4.954]}\end{tabular} & \begin{tabular}[t]{@{}l@{}}0.8883\\{\footnotesize[0.8872--0.8899]}\end{tabular} \\
    Direct & PBE0 & 12 & \begin{tabular}[t]{@{}l@{}}2.417\\{\footnotesize[2.417--2.417]}\end{tabular} & \begin{tabular}[t]{@{}l@{}}0.2821\\{\footnotesize[0.2821--0.2883]}\end{tabular} & \begin{tabular}[t]{@{}l@{}}6.587\\{\footnotesize[6.586--6.631]}\end{tabular} & \begin{tabular}[t]{@{}l@{}}1.214\\{\footnotesize[1.213--1.219]}\end{tabular} \\
    Direct & PBE0 & 48 & \begin{tabular}[t]{@{}l@{}}7.432\\{\footnotesize[7.417--7.478]}\end{tabular} & \begin{tabular}[t]{@{}l@{}}1.696\\{\footnotesize[1.693--1.699]}\end{tabular} & \begin{tabular}[t]{@{}l@{}}11.74\\{\footnotesize[11.73--11.75]}\end{tabular} & \begin{tabular}[t]{@{}l@{}}3.079\\{\footnotesize[3.077--3.08]}\end{tabular} \\
    Direct & PBE0 & 96 & \begin{tabular}[t]{@{}l@{}}15.81\\{\footnotesize[15.73--15.81]}\end{tabular} & \begin{tabular}[t]{@{}l@{}}5.08\\{\footnotesize[5.069--5.081]}\end{tabular} & \begin{tabular}[t]{@{}l@{}}19.7\\{\footnotesize[19.68--19.72]}\end{tabular} & \begin{tabular}[t]{@{}l@{}}6.869\\{\footnotesize[6.869--6.87]}\end{tabular} \\
    Direct & B3LYP & 3 & \begin{tabular}[t]{@{}l@{}}1.419\\{\footnotesize[1.416--1.421]}\end{tabular} & \begin{tabular}[t]{@{}l@{}}0.06778\\{\footnotesize[0.06739--0.06807]}\end{tabular} & \begin{tabular}[t]{@{}l@{}}4.948\\{\footnotesize[4.941--4.957]}\end{tabular} & \begin{tabular}[t]{@{}l@{}}0.8902\\{\footnotesize[0.8886--0.8949]}\end{tabular} \\
    Direct & B3LYP & 12 & \begin{tabular}[t]{@{}l@{}}2.442\\{\footnotesize[2.439--2.442]}\end{tabular} & \begin{tabular}[t]{@{}l@{}}0.2883\\{\footnotesize[0.2882--0.2893]}\end{tabular} & \begin{tabular}[t]{@{}l@{}}6.588\\{\footnotesize[6.583--6.588]}\end{tabular} & \begin{tabular}[t]{@{}l@{}}1.216\\{\footnotesize[1.213--1.218]}\end{tabular} \\
    Direct & B3LYP & 48 & \begin{tabular}[t]{@{}l@{}}7.557\\{\footnotesize[7.506--7.585]}\end{tabular} & \begin{tabular}[t]{@{}l@{}}1.716\\{\footnotesize[1.709--1.719]}\end{tabular} & \begin{tabular}[t]{@{}l@{}}11.75\\{\footnotesize[11.74--11.8]}\end{tabular} & \begin{tabular}[t]{@{}l@{}}3.091\\{\footnotesize[3.088--3.099]}\end{tabular} \\
    Direct & B3LYP & 96 & \begin{tabular}[t]{@{}l@{}}15.94\\{\footnotesize[15.93--15.96]}\end{tabular} & \begin{tabular}[t]{@{}l@{}}5.11\\{\footnotesize[5.104--5.119]}\end{tabular} & \begin{tabular}[t]{@{}l@{}}20.89\\{\footnotesize[20.87--20.89]}\end{tabular} & \begin{tabular}[t]{@{}l@{}}6.85\\{\footnotesize[6.849--6.853]}\end{tabular} \\
    Direct & r$^2$SCAN & 3 & \begin{tabular}[t]{@{}l@{}}1.237\\{\footnotesize[1.236--1.239]}\end{tabular} & \begin{tabular}[t]{@{}l@{}}0.06368\\{\footnotesize[0.06348--0.06369]}\end{tabular} & \begin{tabular}[t]{@{}l@{}}3.237\\{\footnotesize[3.216--3.242]}\end{tabular} & \begin{tabular}[t]{@{}l@{}}0.4622\\{\footnotesize[0.4614--0.4644]}\end{tabular} \\
    Direct & r$^2$SCAN & 12 & \begin{tabular}[t]{@{}l@{}}1.646\\{\footnotesize[1.644--1.649]}\end{tabular} & \begin{tabular}[t]{@{}l@{}}0.2441\\{\footnotesize[0.2435--0.2442]}\end{tabular} & \begin{tabular}[t]{@{}l@{}}4.468\\{\footnotesize[4.452--4.47]}\end{tabular} & \begin{tabular}[t]{@{}l@{}}0.8255\\{\footnotesize[0.8254--0.827]}\end{tabular} \\
    Direct & r$^2$SCAN & 48 & \begin{tabular}[t]{@{}l@{}}5.561\\{\footnotesize[5.557--5.561]}\end{tabular} & \begin{tabular}[t]{@{}l@{}}1.291\\{\footnotesize[1.291--1.291]}\end{tabular} & \begin{tabular}[t]{@{}l@{}}10.22\\{\footnotesize[10.21--10.23]}\end{tabular} & \begin{tabular}[t]{@{}l@{}}3.089\\{\footnotesize[3.083--3.09]}\end{tabular} \\
    Direct & r$^2$SCAN & 96 & \begin{tabular}[t]{@{}l@{}}13.24\\{\footnotesize[13.23--13.24]}\end{tabular} & \begin{tabular}[t]{@{}l@{}}4.18\\{\footnotesize[4.18--4.181]}\end{tabular} & \begin{tabular}[t]{@{}l@{}}18.6\\{\footnotesize[18.6--18.65]}\end{tabular} & \begin{tabular}[t]{@{}l@{}}7.644\\{\footnotesize[7.643--7.644]}\end{tabular} \\
    Direct & r$^2$SCAN-3c & 3 & \begin{tabular}[t]{@{}l@{}}1.247\\{\footnotesize[1.246--1.248]}\end{tabular} & \begin{tabular}[t]{@{}l@{}}0.09323\\{\footnotesize[0.09282--0.09336]}\end{tabular} & \begin{tabular}[t]{@{}l@{}}3.655\\{\footnotesize[3.644--3.662]}\end{tabular} & \begin{tabular}[t]{@{}l@{}}0.7833\\{\footnotesize[0.7807--0.7838]}\end{tabular} \\
    Direct & r$^2$SCAN-3c & 12 & \begin{tabular}[t]{@{}l@{}}2.415\\{\footnotesize[2.413--2.42]}\end{tabular} & \begin{tabular}[t]{@{}l@{}}0.674\\{\footnotesize[0.6733--0.674]}\end{tabular} & \begin{tabular}[t]{@{}l@{}}6.042\\{\footnotesize[6.038--6.058]}\end{tabular} & \begin{tabular}[t]{@{}l@{}}1.422\\{\footnotesize[1.419--1.423]}\end{tabular} \\
    Direct & r$^2$SCAN-3c & 48 & \begin{tabular}[t]{@{}l@{}}8.663\\{\footnotesize[8.66--8.664]}\end{tabular} & \begin{tabular}[t]{@{}l@{}}2.852\\{\footnotesize[2.851--2.855]}\end{tabular} & \begin{tabular}[t]{@{}l@{}}16.53\\{\footnotesize[16.53--16.56]}\end{tabular} & \begin{tabular}[t]{@{}l@{}}5.897\\{\footnotesize[5.897--5.898]}\end{tabular} \\
    Direct & r$^2$SCAN-3c & 96 & \begin{tabular}[t]{@{}l@{}}28.34\\{\footnotesize[28.32--28.36]}\end{tabular} & \begin{tabular}[t]{@{}l@{}}10.18\\{\footnotesize[10.17--10.18]}\end{tabular} & \begin{tabular}[t]{@{}l@{}}32.09\\{\footnotesize[32.09--32.12]}\end{tabular} & \begin{tabular}[t]{@{}l@{}}14.95\\{\footnotesize[14.94--14.95]}\end{tabular} \\
    Direct & M06-2X & 3 & \begin{tabular}[t]{@{}l@{}}1.491\\{\footnotesize[1.49--1.495]}\end{tabular} & \begin{tabular}[t]{@{}l@{}}0.08234\\{\footnotesize[0.0823--0.08242]}\end{tabular} & \begin{tabular}[t]{@{}l@{}}5.051\\{\footnotesize[5.048--5.06]}\end{tabular} & \begin{tabular}[t]{@{}l@{}}0.9166\\{\footnotesize[0.916--0.9171]}\end{tabular} \\
    Direct & M06-2X & 12 & \begin{tabular}[t]{@{}l@{}}2.521\\{\footnotesize[2.517--2.522]}\end{tabular} & \begin{tabular}[t]{@{}l@{}}0.3118\\{\footnotesize[0.3115--0.312]}\end{tabular} & \begin{tabular}[t]{@{}l@{}}7.145\\{\footnotesize[7.137--7.145]}\end{tabular} & \begin{tabular}[t]{@{}l@{}}1.282\\{\footnotesize[1.277--1.283]}\end{tabular} \\
    Direct & M06-2X & 48 & \begin{tabular}[t]{@{}l@{}}9.555\\{\footnotesize[9.53--9.576]}\end{tabular} & \begin{tabular}[t]{@{}l@{}}1.891\\{\footnotesize[1.889--1.898]}\end{tabular} & \begin{tabular}[t]{@{}l@{}}13.23\\{\footnotesize[13.21--13.25]}\end{tabular} & \begin{tabular}[t]{@{}l@{}}3.816\\{\footnotesize[3.814--3.82]}\end{tabular} \\
    Direct & M06-2X & 96 & \begin{tabular}[t]{@{}l@{}}19.55\\{\footnotesize[19.51--19.56]}\end{tabular} & \begin{tabular}[t]{@{}l@{}}5.393\\{\footnotesize[5.38--5.399]}\end{tabular} & \begin{tabular}[t]{@{}l@{}}24.47\\{\footnotesize[24.47--24.5]}\end{tabular} & \begin{tabular}[t]{@{}l@{}}8.742\\{\footnotesize[8.74--8.746]}\end{tabular} \\
    Direct & $\omega$B97M-V & 3 & \begin{tabular}[t]{@{}l@{}}2.694\\{\footnotesize[2.686--2.697]}\end{tabular} & \begin{tabular}[t]{@{}l@{}}0.2919\\{\footnotesize[0.2918--0.2922]}\end{tabular} & \begin{tabular}[t]{@{}l@{}}6.88\\{\footnotesize[6.855--6.892]}\end{tabular} & \begin{tabular}[t]{@{}l@{}}1.637\\{\footnotesize[1.633--1.637]}\end{tabular} \\
    Direct & $\omega$B97M-V & 12 & \begin{tabular}[t]{@{}l@{}}4.242\\{\footnotesize[4.225--4.249]}\end{tabular} & \begin{tabular}[t]{@{}l@{}}0.6447\\{\footnotesize[0.6398--0.6618]}\end{tabular} & \begin{tabular}[t]{@{}l@{}}15.22\\{\footnotesize[15.22--15.22]}\end{tabular} & \begin{tabular}[t]{@{}l@{}}3.985\\{\footnotesize[3.983--3.99]}\end{tabular} \\
    Direct & $\omega$B97M-V & 48 & \begin{tabular}[t]{@{}l@{}}48.07\\{\footnotesize[48.07--48.1]}\end{tabular} & \begin{tabular}[t]{@{}l@{}}5.268\\{\footnotesize[5.243--5.277]}\end{tabular} & \begin{tabular}[t]{@{}l@{}}95.95\\{\footnotesize[95.93--95.96]}\end{tabular} & \begin{tabular}[t]{@{}l@{}}32.41\\{\footnotesize[32.41--32.41]}\end{tabular} \\
    Direct & $\omega$B97M-V & 96 & \begin{tabular}[t]{@{}l@{}}166.5\\{\footnotesize[166.5--166.6]}\end{tabular} & \begin{tabular}[t]{@{}l@{}}12.65\\{\footnotesize[12.65--12.66]}\end{tabular} & \begin{tabular}[t]{@{}l@{}}369.1\\{\footnotesize[369.1--369.1]}\end{tabular} & \begin{tabular}[t]{@{}l@{}}119.1\\{\footnotesize[119.1--119.1]}\end{tabular} \\
    DF & HF & 3 & \begin{tabular}[t]{@{}l@{}}0.6351\\{\footnotesize[0.633--0.636]}\end{tabular} & \begin{tabular}[t]{@{}l@{}}0.01873\\{\footnotesize[0.01873--0.01883]}\end{tabular} & \begin{tabular}[t]{@{}l@{}}2.554\\{\footnotesize[2.538--2.563]}\end{tabular} & \begin{tabular}[t]{@{}l@{}}0.2713\\{\footnotesize[0.2695--0.2722]}\end{tabular} \\
    DF & HF & 12 & \begin{tabular}[t]{@{}l@{}}0.8528\\{\footnotesize[0.8519--0.8531]}\end{tabular} & \begin{tabular}[t]{@{}l@{}}0.1062\\{\footnotesize[0.1062--0.1064]}\end{tabular} & \begin{tabular}[t]{@{}l@{}}2.606\\{\footnotesize[2.593--2.613]}\end{tabular} & \begin{tabular}[t]{@{}l@{}}0.3043\\{\footnotesize[0.3042--0.3049]}\end{tabular} \\
    DF & HF & 48 & \begin{tabular}[t]{@{}l@{}}2.645\\{\footnotesize[2.642--2.647]}\end{tabular} & \begin{tabular}[t]{@{}l@{}}0.6359\\{\footnotesize[0.6352--0.6364]}\end{tabular} & \begin{tabular}[t]{@{}l@{}}3.943\\{\footnotesize[3.936--3.949]}\end{tabular} & \begin{tabular}[t]{@{}l@{}}1.026\\{\footnotesize[1.025--1.026]}\end{tabular} \\
    DF & HF & 96 & \begin{tabular}[t]{@{}l@{}}16.34\\{\footnotesize[16.34--16.35]}\end{tabular} & \begin{tabular}[t]{@{}l@{}}3.781\\{\footnotesize[3.781--3.783]}\end{tabular} & \begin{tabular}[t]{@{}l@{}}16.53\\{\footnotesize[16.52--16.53]}\end{tabular} & \begin{tabular}[t]{@{}l@{}}6.907\\{\footnotesize[6.905--6.908]}\end{tabular} \\
    DF & PBE & 3 & \begin{tabular}[t]{@{}l@{}}0.8566\\{\footnotesize[0.8559--0.8568]}\end{tabular} & \begin{tabular}[t]{@{}l@{}}0.04187\\{\footnotesize[0.04187--0.04201]}\end{tabular} & \begin{tabular}[t]{@{}l@{}}3.059\\{\footnotesize[3.014--3.06]}\end{tabular} & \begin{tabular}[t]{@{}l@{}}0.2942\\{\footnotesize[0.2937--0.2956]}\end{tabular} \\
    DF & PBE & 12 & \begin{tabular}[t]{@{}l@{}}1.228\\{\footnotesize[1.227--1.23]}\end{tabular} & \begin{tabular}[t]{@{}l@{}}0.1611\\{\footnotesize[0.161--0.1612]}\end{tabular} & \begin{tabular}[t]{@{}l@{}}3.727\\{\footnotesize[3.723--3.74]}\end{tabular} & \begin{tabular}[t]{@{}l@{}}0.4595\\{\footnotesize[0.4589--0.4601]}\end{tabular} \\
    DF & PBE & 48 & \begin{tabular}[t]{@{}l@{}}4.118\\{\footnotesize[4.117--4.119]}\end{tabular} & \begin{tabular}[t]{@{}l@{}}1.086\\{\footnotesize[1.086--1.087]}\end{tabular} & \begin{tabular}[t]{@{}l@{}}6.373\\{\footnotesize[6.37--6.377]}\end{tabular} & \begin{tabular}[t]{@{}l@{}}1.477\\{\footnotesize[1.477--1.479]}\end{tabular} \\
    DF & PBE & 96 & \begin{tabular}[t]{@{}l@{}}9.86\\{\footnotesize[9.854--9.863]}\end{tabular} & \begin{tabular}[t]{@{}l@{}}3.58\\{\footnotesize[3.58--3.581]}\end{tabular} & \begin{tabular}[t]{@{}l@{}}11.84\\{\footnotesize[11.84--11.84]}\end{tabular} & \begin{tabular}[t]{@{}l@{}}4.239\\{\footnotesize[4.237--4.243]}\end{tabular} \\
    DF & PBE0 & 3 & \begin{tabular}[t]{@{}l@{}}0.8712\\{\footnotesize[0.8677--0.8727]}\end{tabular} & \begin{tabular}[t]{@{}l@{}}0.04558\\{\footnotesize[0.04538--0.04559]}\end{tabular} & \begin{tabular}[t]{@{}l@{}}3.004\\{\footnotesize[2.996--3.005]}\end{tabular} & \begin{tabular}[t]{@{}l@{}}0.3502\\{\footnotesize[0.3498--0.3527]}\end{tabular} \\
    DF & PBE0 & 12 & \begin{tabular}[t]{@{}l@{}}1.263\\{\footnotesize[1.262--1.265]}\end{tabular} & \begin{tabular}[t]{@{}l@{}}0.1899\\{\footnotesize[0.1896--0.1906]}\end{tabular} & \begin{tabular}[t]{@{}l@{}}3.476\\{\footnotesize[3.476--3.495]}\end{tabular} & \begin{tabular}[t]{@{}l@{}}0.5223\\{\footnotesize[0.5222--0.5247]}\end{tabular} \\
    DF & PBE0 & 48 & \begin{tabular}[t]{@{}l@{}}4.763\\{\footnotesize[4.762--4.763]}\end{tabular} & \begin{tabular}[t]{@{}l@{}}1.298\\{\footnotesize[1.298--1.3]}\end{tabular} & \begin{tabular}[t]{@{}l@{}}6.683\\{\footnotesize[6.67--6.688]}\end{tabular} & \begin{tabular}[t]{@{}l@{}}2.078\\{\footnotesize[2.077--2.079]}\end{tabular} \\
    DF & PBE0 & 96 & \begin{tabular}[t]{@{}l@{}}20.09\\{\footnotesize[20.08--20.13]}\end{tabular} & \begin{tabular}[t]{@{}l@{}}5.808\\{\footnotesize[5.807--5.811]}\end{tabular} & \begin{tabular}[t]{@{}l@{}}22.78\\{\footnotesize[22.78--22.78]}\end{tabular} & \begin{tabular}[t]{@{}l@{}}10.37\\{\footnotesize[10.37--10.37]}\end{tabular} \\
    DF & B3LYP & 3 & \begin{tabular}[t]{@{}l@{}}0.8797\\{\footnotesize[0.8788--0.881]}\end{tabular} & \begin{tabular}[t]{@{}l@{}}0.05017\\{\footnotesize[0.04999--0.05031]}\end{tabular} & \begin{tabular}[t]{@{}l@{}}2.995\\{\footnotesize[2.985--2.997]}\end{tabular} & \begin{tabular}[t]{@{}l@{}}0.3533\\{\footnotesize[0.3529--0.354]}\end{tabular} \\
    DF & B3LYP & 12 & \begin{tabular}[t]{@{}l@{}}1.289\\{\footnotesize[1.289--1.291]}\end{tabular} & \begin{tabular}[t]{@{}l@{}}0.2051\\{\footnotesize[0.2048--0.2054]}\end{tabular} & \begin{tabular}[t]{@{}l@{}}3.501\\{\footnotesize[3.475--3.512]}\end{tabular} & \begin{tabular}[t]{@{}l@{}}0.5247\\{\footnotesize[0.5247--0.5269]}\end{tabular} \\
    DF & B3LYP & 48 & \begin{tabular}[t]{@{}l@{}}4.809\\{\footnotesize[4.808--4.811]}\end{tabular} & \begin{tabular}[t]{@{}l@{}}1.317\\{\footnotesize[1.317--1.317]}\end{tabular} & \begin{tabular}[t]{@{}l@{}}6.69\\{\footnotesize[6.673--6.693]}\end{tabular} & \begin{tabular}[t]{@{}l@{}}2.084\\{\footnotesize[2.081--2.085]}\end{tabular} \\
    DF & B3LYP & 96 & \begin{tabular}[t]{@{}l@{}}20.18\\{\footnotesize[20.18--20.2]}\end{tabular} & \begin{tabular}[t]{@{}l@{}}5.844\\{\footnotesize[5.844--5.848]}\end{tabular} & \begin{tabular}[t]{@{}l@{}}23.98\\{\footnotesize[23.98--23.99]}\end{tabular} & \begin{tabular}[t]{@{}l@{}}10.38\\{\footnotesize[10.38--10.38]}\end{tabular} \\
    DF & r$^2$SCAN & 3 & \begin{tabular}[t]{@{}l@{}}0.979\\{\footnotesize[0.9769--0.9806]}\end{tabular} & \begin{tabular}[t]{@{}l@{}}0.0715\\{\footnotesize[0.07145--0.07161]}\end{tabular} & \begin{tabular}[t]{@{}l@{}}3.072\\{\footnotesize[3.067--3.076]}\end{tabular} & \begin{tabular}[t]{@{}l@{}}0.3184\\{\footnotesize[0.318--0.3189]}\end{tabular} \\
    DF & r$^2$SCAN & 12 & \begin{tabular}[t]{@{}l@{}}1.419\\{\footnotesize[1.418--1.422]}\end{tabular} & \begin{tabular}[t]{@{}l@{}}0.2359\\{\footnotesize[0.2357--0.236]}\end{tabular} & \begin{tabular}[t]{@{}l@{}}3.709\\{\footnotesize[3.705--3.721]}\end{tabular} & \begin{tabular}[t]{@{}l@{}}0.5287\\{\footnotesize[0.5287--0.5292]}\end{tabular} \\
    DF & r$^2$SCAN & 48 & \begin{tabular}[t]{@{}l@{}}5.271\\{\footnotesize[5.268--5.281]}\end{tabular} & \begin{tabular}[t]{@{}l@{}}1.344\\{\footnotesize[1.343--1.344]}\end{tabular} & \begin{tabular}[t]{@{}l@{}}7.726\\{\footnotesize[7.712--7.731]}\end{tabular} & \begin{tabular}[t]{@{}l@{}}2.207\\{\footnotesize[2.205--2.21]}\end{tabular} \\
    DF & r$^2$SCAN & 96 & \begin{tabular}[t]{@{}l@{}}12.42\\{\footnotesize[12.41--12.42]}\end{tabular} & \begin{tabular}[t]{@{}l@{}}4.033\\{\footnotesize[4.031--4.034]}\end{tabular} & \begin{tabular}[t]{@{}l@{}}15.06\\{\footnotesize[15.05--15.07]}\end{tabular} & \begin{tabular}[t]{@{}l@{}}6.14\\{\footnotesize[6.137--6.145]}\end{tabular} \\
    DF & r$^2$SCAN-3c & 3 & \begin{tabular}[t]{@{}l@{}}0.9411\\{\footnotesize[0.9405--0.9427]}\end{tabular} & \begin{tabular}[t]{@{}l@{}}0.07688\\{\footnotesize[0.07662--0.07695]}\end{tabular} & \begin{tabular}[t]{@{}l@{}}3.165\\{\footnotesize[3.157--3.182]}\end{tabular} & \begin{tabular}[t]{@{}l@{}}0.352\\{\footnotesize[0.3501--0.3526]}\end{tabular} \\
    DF & r$^2$SCAN-3c & 12 & \begin{tabular}[t]{@{}l@{}}1.889\\{\footnotesize[1.886--1.895]}\end{tabular} & \begin{tabular}[t]{@{}l@{}}0.3092\\{\footnotesize[0.3084--0.3097]}\end{tabular} & \begin{tabular}[t]{@{}l@{}}3.963\\{\footnotesize[3.961--3.965]}\end{tabular} & \begin{tabular}[t]{@{}l@{}}0.6042\\{\footnotesize[0.6022--0.6062]}\end{tabular} \\
    DF & r$^2$SCAN-3c & 48 & \begin{tabular}[t]{@{}l@{}}7.481\\{\footnotesize[7.48--7.486]}\end{tabular} & \begin{tabular}[t]{@{}l@{}}2.349\\{\footnotesize[2.347--2.354]}\end{tabular} & \begin{tabular}[t]{@{}l@{}}10.02\\{\footnotesize[10.01--10.02]}\end{tabular} & \begin{tabular}[t]{@{}l@{}}3.45\\{\footnotesize[3.449--3.451]}\end{tabular} \\
    DF & r$^2$SCAN-3c & 96 & \begin{tabular}[t]{@{}l@{}}23.05\\{\footnotesize[23.04--23.06]}\end{tabular} & \begin{tabular}[t]{@{}l@{}}6.594\\{\footnotesize[6.593--6.594]}\end{tabular} & \begin{tabular}[t]{@{}l@{}}23.12\\{\footnotesize[23.12--23.14]}\end{tabular} & \begin{tabular}[t]{@{}l@{}}10.07\\{\footnotesize[10.06--10.07]}\end{tabular} \\
    DF & M06-2X & 3 & \begin{tabular}[t]{@{}l@{}}0.9495\\{\footnotesize[0.9494--0.9507]}\end{tabular} & \begin{tabular}[t]{@{}l@{}}0.06488\\{\footnotesize[0.06474--0.06492]}\end{tabular} & \begin{tabular}[t]{@{}l@{}}3.056\\{\footnotesize[3.056--3.059]}\end{tabular} & \begin{tabular}[t]{@{}l@{}}0.3773\\{\footnotesize[0.376--0.3774]}\end{tabular} \\
    DF & M06-2X & 12 & \begin{tabular}[t]{@{}l@{}}1.387\\{\footnotesize[1.386--1.391]}\end{tabular} & \begin{tabular}[t]{@{}l@{}}0.2347\\{\footnotesize[0.234--0.2349]}\end{tabular} & \begin{tabular}[t]{@{}l@{}}3.736\\{\footnotesize[3.721--3.738]}\end{tabular} & \begin{tabular}[t]{@{}l@{}}0.5955\\{\footnotesize[0.595--0.5978]}\end{tabular} \\
    DF & M06-2X & 48 & \begin{tabular}[t]{@{}l@{}}6.343\\{\footnotesize[6.343--6.348]}\end{tabular} & \begin{tabular}[t]{@{}l@{}}1.497\\{\footnotesize[1.496--1.498]}\end{tabular} & \begin{tabular}[t]{@{}l@{}}8.167\\{\footnotesize[8.159--8.184]}\end{tabular} & \begin{tabular}[t]{@{}l@{}}2.809\\{\footnotesize[2.808--2.81]}\end{tabular} \\
    DF & M06-2X & 96 & \begin{tabular}[t]{@{}l@{}}23.32\\{\footnotesize[23.31--23.32]}\end{tabular} & \begin{tabular}[t]{@{}l@{}}6.138\\{\footnotesize[6.137--6.141]}\end{tabular} & \begin{tabular}[t]{@{}l@{}}27.65\\{\footnotesize[27.65--27.66]}\end{tabular} & \begin{tabular}[t]{@{}l@{}}12.27\\{\footnotesize[12.27--12.28]}\end{tabular} \\
    DF & $\omega$B97M-V & 3 & \begin{tabular}[t]{@{}l@{}}1.488\\{\footnotesize[1.486--1.49]}\end{tabular} & \begin{tabular}[t]{@{}l@{}}0.2621\\{\footnotesize[0.2616--0.2623]}\end{tabular} & \begin{tabular}[t]{@{}l@{}}3.838\\{\footnotesize[3.837--3.87]}\end{tabular} & \begin{tabular}[t]{@{}l@{}}0.7513\\{\footnotesize[0.7488--0.7514]}\end{tabular} \\
    DF & $\omega$B97M-V & 12 & \begin{tabular}[t]{@{}l@{}}4.619\\{\footnotesize[4.61--4.622]}\end{tabular} & \begin{tabular}[t]{@{}l@{}}0.6252\\{\footnotesize[0.6246--0.6299]}\end{tabular} & \begin{tabular}[t]{@{}l@{}}10.4\\{\footnotesize[10.39--10.46]}\end{tabular} & \begin{tabular}[t]{@{}l@{}}2.792\\{\footnotesize[2.787--2.793]}\end{tabular} \\
    DF & $\omega$B97M-V & 48 & \begin{tabular}[t]{@{}l@{}}43.78\\{\footnotesize[43.78--43.79]}\end{tabular} & \begin{tabular}[t]{@{}l@{}}5.62\\{\footnotesize[5.618--5.62]}\end{tabular} & \begin{tabular}[t]{@{}l@{}}88.95\\{\footnotesize[88.95--88.95]}\end{tabular} & \begin{tabular}[t]{@{}l@{}}30.83\\{\footnotesize[30.83--30.84]}\end{tabular} \\
    DF & $\omega$B97M-V & 96 & \begin{tabular}[t]{@{}l@{}}188.7\\{\footnotesize[188.7--188.9]}\end{tabular} & \begin{tabular}[t]{@{}l@{}}22.78\\{\footnotesize[22.78--22.78]}\end{tabular} & \begin{tabular}[t]{@{}l@{}}376.2\\{\footnotesize[376.2--376.2]}\end{tabular} & \begin{tabular}[t]{@{}l@{}}125.9\\{\footnotesize[125.9--125.9]}\end{tabular} \\
}

%% file: sections/abstract.tex
GenerativeQC is an open-source, GPU-accelerated quantum chemistry program that computes
energies and analytic nuclear forces from Hartree--Fock (HF) and density
functional theory (DFT) to closed-shell density-fitted CCSD(T) and an
MP2-assisted CCSD(T)/\allowbreak{}complete-basis-set (CBS) composite. Its
scientific compiler translates symbolic equations into C++ and CUDA code and
uses automatic differentiation to generate the derivative and response
programs required for analytic forces. A native runtime executes these
programs together with hand-written iterative solvers and integral routines.
Across Direct and density-fitted HF and DFT benchmarks for eight methods on
water clusters of up to 96 atoms, GenerativeQC is faster than GPU4PySCF in
every tested case on both RTX~5090 and H100 GPUs. For correlated wave
functions, compiler-generated derivative contractions and transpose Jacobian
operators enable complete density-fitted CCSD(T) and CCSD(T)/CBS analytic
forces on GPUs; a CCSD(T)/CBS energy-and-force calculation for heptane (23
atoms) takes \GQC{cbs.h100.last.min}~min on a single H100 GPU and
\GQC{cbs.rtx5090.last.min}~min on an RTX~5090.

%% file: sections/intro.tex
\section{Introduction}
\label{sec:intro}

Quantum chemistry programs must offer a broad range of methods, from
Hartree--Fock (HF) and density functional theory (DFT) to correlated
wave-function theories, while running them efficiently on diverse and rapidly
changing hardware.\cite{Smith_JChemPhys_2020_v152_p184108}
Analytic nuclear forces enable geometry optimization, vibrational analysis, and
molecular dynamics,\cite{Bozkaya_JChemPhys_2026_v165_p162501} yet computing
them remains among the most demanding tasks in quantum
chemistry.\cite{Stocks_JChemTheoryComput_2024_v20_p7503} Programs for graphics
processing units (GPUs) now provide HF and DFT energies and nuclear
gradients:\cite{Seritan_WiresComputMolSci_2021_v11_pe1494,Li_JPhysChemA_2025_v129_p1459,Wu_WiresComputMolSci_2025_v15_pe70008,Stocks_JChemTheoryComput_2024_v20_p7503}
GPU kernels evaluate the electron-repulsion integrals, build the Coulomb and
exchange matrices, and compute the exchange--correlation potentials and their
derivatives.\cite{Seritan_WiresComputMolSci_2021_v11_pe1494,Li_JPhysChemA_2025_v129_p1459,Wu_WiresComputMolSci_2025_v15_pe70008}
Beyond mean-field theory, GPU implementations provide resolution-of-the-identity MP2
gradients\cite{Stocks_JChemTheoryComput_2024_v20_p2505} and correlated methods
up to coupled cluster with single, double and perturbative triple excitations,
CCSD(T).\cite{Raghavachari_ChemPhysLett_1989_v157_p479,Guo_WiresComputMolSci_2025_v15_pe70034,Datta_JChemTheoryComput_2023_v19_p7640,Fajen_JPhysChemA_2026_v130_p2225}
Performance gains on GPUs, however, usually require redesigned algorithms and
kernels tuned for each
architecture.\cite{Li_JPhysChemA_2025_v129_p1459,Wu_WiresComputMolSci_2025_v15_pe70008}

Hardware tuning is only part of the challenge: the algebraic complexity of
quantum chemistry methods also makes analytic forces difficult to
implement,\cite{Smith_JChemPhys_2020_v152_p184108} especially for
nonvariational energies that require response equations alongside explicit
energy derivatives.\cite{Handy_JChemPhys_1984_v81_p5031} Code generation
addresses this complexity by deriving executable programs directly from
symbolic equations, avoiding laborious and error-prone manual
implementation.\cite{MacLeod_JChemPhys_2015_v142_p51103} For example, the
Tensor Contraction Engine translated many-body equations into parallel
code,\cite{Hirata_JPhysChemA_2003_v107_p9887} and SMITH3 generated equations
and code for the nuclear gradients of fully internally contracted CASPT2, whose
complexity had long hampered
implementation.\cite{MacLeod_JChemPhys_2015_v142_p51103} Automatic
differentiation (AD) propagates derivatives through elementary program
operations via the chain rule, yielding exact derivatives of arbitrary order
without finite differences or hand-derived
formulas.\cite{Zhang_JChemPhys_2022_v157_p204801,Abbott_JPhysChemLett_2021_v12_p3232}
Implicit differentiation of the converged solutions of iterative solvers
supplies the response terms without an explicit implementation of the
coupled-perturbed HF and coupled-cluster $\Lambda$
equations.\cite{Zhang_JChemPhys_2022_v157_p204801} Quax obtained CCSD(T)
nuclear derivatives by AD,\cite{Abbott_JPhysChemLett_2021_v12_p3232} and
PySCFAD, which builds on the JAX array library, demonstrated CCSD gradients and
subsequently density-fitted canonical and local natural orbital CCSD(T)
gradients.\cite{Zhang_JChemPhys_2022_v157_p204801,Zhang_JChemPhys_2024_v161_p014109}

Combining derivative automation with specialized GPU kernels is particularly
demanding for correlated wave-function forces. Density-fitted (DF) CCSD(T)
gradients\cite{Bozkaya_JChemPhys_2026_v165_p162501} and
CCSD(T)/\allowbreak{}complete-basis-set (CBS) composite
gradients\cite{Warden_JChemPhys_2020_v152_p124109} are available on CPUs,
whereas GPU implementations have addressed individual parts of the problem:
evaluating the (T) gradient contribution on GPUs while solving the $\Lambda$
equations on CPUs,\cite{Guo_MolPhys_2021_v119_pe1974591} deriving CCSD response
operations from a contraction graph,\cite{Guerrero_arXiv_2026_p2607.05622}
solving the $\Lambda$ equations for triples corrections and density
matrices,\cite{Datta_JChemPhys_2026_v165_p124103,Pototschnig_JChemTheoryComput_2021_v17_p5509}
or applying CBS-extrapolated CCSD(T) to interaction
energies.\cite{Fajen_JPhysChemA_2026_v130_p2225}
GenerativeQC instead centers on a domain-specific scientific compiler. Starting
from symbolic descriptions of integrals, tensor equations, and methods, it
applies compile-time AD to derive the derivative and response programs for
analytic forces, then lowers both energy and derivative programs to native C++
and CUDA kernels that run alongside hand-written solvers and integral routines.
In DF-CCSD(T), AD generates the transpose action of the CCSD residual Jacobian
as a GPU operator that native GMRES solvers call to solve the $\Lambda$
equations, enabling complete analytic forces for DF-CCSD(T) and the
CCSD(T)/\allowbreak{}CBS composite on GPUs
(Section~\ref{sec:methods-dfccsdt}).
To achieve high performance in practice, coding agents iteratively refine the
compiler lowering rules and runtime routines using timing profiles and work
counts, adopting candidate changes as defaults only after verifying agreement
with independent references and shortening complete energy-and-force calls
(Section~\ref{sec:methods-optimization}).

Section~\ref{sec:methods} describes the supported electronic-structure methods,
the scientific compiler, the DF-CCSD(T) force calculation, and the agent-driven
optimization loop. Section~\ref{sec:results} compares the HF and DFT paths on
RTX~5090 and H100 GPUs with GPU4PySCF, an open-source GPU extension of PySCF
that provides the same methods with analytic
forces,\cite{Li_JPhysChemA_2025_v129_p1459,Wu_WiresComputMolSci_2025_v15_pe70008}
and evaluates the accuracy and computational cost of the
CCSD(T)/\allowbreak{}CBS composite for $n$-alkanes up to heptane.

%% file: sections/methods.tex
\section{Methods and implementation}
\label{sec:methods}

GenerativeQC separates mathematical method definition from hardware execution
via a Python scientific compiler and a native C++/CUDA runtime.

\subsection{Quantum-chemical methods and analytic forces}
\label{sec:methods-scope}

The program computes energies and analytic nuclear forces on CUDA GPUs for HF,
Kohn--Sham DFT, second-order M\o{}ller--Plesset perturbation theory (MP2),
CCSD(T),\cite{Raghavachari_ChemPhysLett_1989_v157_p479} and an MP2-assisted
CCSD(T)/\allowbreak{}CBS composite.

\paragraph{Hartree--Fock and Kohn--Sham DFT.}
Restricted and unrestricted HF and Kohn--Sham calculations use one of two
treatments of the two-electron interaction. In the \emph{Direct} treatment,
the Coulomb (J) and exchange (K) matrices are built from four-center
electron-repulsion integrals. In the \emph{density-fitted} (DF) treatment, J
and K are built from three-center integrals over an auxiliary basis and the
two-center Coulomb metric, which is factorized by an eigendecomposition with a
relative cutoff of $10^{-10}$ times its largest eigenvalue. DF forces are
evaluated consistently with the DF energy, including the nuclear derivatives of
the auxiliary basis functions and the Coulomb metric. The functionals
benchmarked in this work range
from semilocal approximations such as PBE\cite{Perdew_PhysRevLett_1996_v77_p3865}
and r$^2$SCAN\cite{Furness_JPhysChemLett_2020_v11_p8208} to the global hybrids
PBE0\cite{Adamo_JChemPhys_1999_v110_p6158} and
B3LYP,\cite{Becke_JChemPhys_1993_v98_p5648,Stephens_JPhysChem_1994_v98_p11623}
the Minnesota hybrid M06-2X\cite{Zhao_TheorChemAcc_2008_v120_p215} and the
range-separated hybrid meta-GGA
$\omega$B97M-V,\cite{Mardirossian_JChemPhys_2016_v144_p214110} whose VV10
nonlocal correlation\cite{Vydrov_JChemPhys_2010_v133_p244103} is evaluated
self-consistently. The
r$^2$SCAN-3c composite\cite{Grimme_JChemPhys_2021_v154_p64103} combines
r$^2$SCAN in its def2-mTZVPP basis with D4
dispersion\cite{Caldeweyher_JChemPhys_2019_v150_p154122} and gCP
corrections,\cite{Kruse_JChemPhys_2012_v136_p154101} which enter both the
energy and the forces.
For B3LYP, r$^2$SCAN, M06-2X, and the semilocal part of $\omega$B97M-V, the
compiler generates the exchange--correlation point code from one expression
graph per functional translated at build time from Libxc~7.0.0
definitions,\cite{Lehtola_Softwarex_2018_v7_p1} differentiating this graph to
produce both the SCF potential and force point derivatives without runtime
calls to Libxc. PBE and PBE0 use a hand-written point evaluator.

\paragraph{MP2 and coupled-cluster theory.}
Closed-shell MP2 and CCSD(T) energies and analytic forces are available with
four-center integrals (for CCSD(T) in small systems) and with density-fitted
correlation. Both DF-MP2
and DF-CCSD(T) use one of two fitting models. In the \emph{correlation-only DF} model, a conventional
all-electron RHF calculation with four-center integrals supplies the reference
orbitals and Fock matrix, the orbital response uses the same exact integrals,
and only the two-electron integrals of the correlation treatment are fitted.
In the \emph{full-DF} model, the reference is a stationary RI-JK RHF
solution and the same RI operator acts in the orbital response. One
auxiliary basis serves both the JK and the correlation fitting, and the
analytic forces include the three-center, auxiliary-metric, nuclear-center,
orbital and Pulay derivatives. Both models treat closed-shell singlets with all
electrons correlated and follow the analytic-gradient theory of
CCSD(T)\cite{Scuseria_JChemPhys_1991_v94_p442,Lee_JChemPhys_1991_v94_p6229}
and its density-fitted formulation by Bozkaya and
Sherrill.\cite{Bozkaya_JChemPhys_2017_v147_p44104}

\paragraph{CCSD(T)/CBS composite.}
The composite energy adds a CCSD(T)$-$MP2 correction in a smaller basis $S$
to HF and MP2 energies extrapolated from two basis sets with cardinal numbers
$X<Y$:
\begin{equation}
  E = E_\mathrm{HF}^{\infty} + E_\mathrm{corr,MP2}^{\infty}
    + \left(E_\mathrm{CCSD(T)}^{S} - E_\mathrm{MP2}^{S}\right).
  \label{eq:methods-cbs-energy}
\end{equation}
This additive form is the established composite, or focal-point, approach in
which a higher-order correction from a small basis is added to a lower-order
CBS
estimate.\cite{East_JChemPhys_1993_v99_p4638,Jurecka_ChemPhysLett_2002_v365_p89,Karton_WiresComputMolSci_2016_v6_p292,Warden_JChemPhys_2020_v152_p124109}
The MP2 correlation energy is extrapolated with
$E_\mathrm{corr}(X) = E_\mathrm{corr}^{\infty} + A_\mathrm{corr} X^{-p}$ and
$p=3$,\cite{Helgaker_JChemPhys_1997_v106_p9639,Halkier_ChemPhysLett_1998_v286_p243}
which gives the weights $w_X = -X^p/(Y^p-X^p)$ and $w_Y = Y^p/(Y^p-X^p)$.
The HF energy is extrapolated separately with
$E_\mathrm{HF}(X) = E_\mathrm{HF}^{\infty} + B_\mathrm{HF} e^{-\alpha X}$ and the fixed
global exponent $\alpha = 1.63$ of Halkier and
co-workers.\cite{Halkier_ChemPhysLett_1999_v302_p437} Here $X=2$
and $Y=3$ (cc-pVDZ and cc-pVTZ\cite{Dunning_JChemPhys_1989_v90_p1007}), and
$S$ is cc-pVDZ, so the correction is the difference between DF-CCSD(T) and
DF-MP2 in cc-pVDZ. In the full-DF composite benchmarked below, the
cc-pV$X$Z-JKFIT auxiliary sets\cite{Weigend_PhysChemChemPhys_2002_v4_p4285}
fit both the JK and the correlation terms of each basis group.

Because the weights do not depend on the geometry, the analytic force is the
same linear combination of component forces,
\begin{equation}
  \mathbf F = \mathbf F_\mathrm{HF}^{\infty}
    + \sum_{K=X,Y} w_K \left(\mathbf F_\mathrm{MP2}^{K}
    - \mathbf F_\mathrm{HF}^{K}\right)
    + \mathbf F_\mathrm{CCSD(T)}^{S} - \mathbf F_\mathrm{MP2}^{S},
  \label{eq:methods-cbs-force}
\end{equation}
where $\mathbf F_\mathrm{HF}^{\infty}$ combines the HF forces with the HF
weights and the MP2 terms are total-energy forces; $p$ and $\alpha$ are fixed
inputs. Composite gradients of this
type are available on CPUs, for example through Psi4's \texttt{cbs()}
driver;\cite{Smith_JChemPhys_2020_v152_p184108,Warden_JChemPhys_2020_v152_p124109}
in GenerativeQC, the operator actions and tensor contractions of every
component execute on the GPU (with host-side Krylov control,
Section~\ref{sec:methods-dfccsdt}). The components run in two basis groups, cc-pVDZ (HF, MP2 and
CCSD(T)) and cc-pVTZ (HF and MP2), one after the other. Within a group, the
components share one converged canonical RHF reference and its orbitals, and
the RI-JK plan with its metric and factors stays resident across the MP2 and
CCSD(T) response steps. Because the correction basis is the lower MP2 basis, the cc-pVDZ
MP2 result is computed once and reused.

\paragraph{Further capabilities.}
GenerativeQC also provides D3\cite{Grimme_JChemPhys_2010_v132_p154104} and
D4\cite{Caldeweyher_JChemPhys_2019_v150_p154122} dispersion corrections,
scalar effective core potentials, the GFN2-xTB tight-binding
method,\cite{Bannwarth_JChemTheoryComput_2019_v15_p1652} analytic Hessians for
selected functionals and a native geometry optimizer. The same implementations are accessible through Python, C, and C++ interfaces
as well as a native command-line program.

\subsection{Scientific compiler}
\label{sec:methods-compiler}

The scientific compiler organizes code generation around three intermediate
representations (IRs) and an offline lowering pipeline.

\paragraph{Representations and differentiation.}
These three IRs separate domain mathematics from device scheduling. IntegralIR
encodes Gaussian shell structure, recurrence relations and nuclear derivative
centers without warp, block or shared-memory parameters; for example,
one-electron overlap, kinetic and nuclear-attraction primitives are
differentiated once symbolically and lowered to shell-warp CUDA kernels.
TensorIR represents real tensor equations over typed occupied, virtual,
atomic-orbital and auxiliary index spaces with explicit permutation symmetries
(for example, the simultaneous pair exchange $t_{ij}^{ab}=t_{ji}^{ba}$ of
restricted spatial amplitudes) and exact rational coefficients. MethodIR
composes complete electronic-structure workflows from these building blocks.
TensorIR defines forward Jacobian--vector products (JVPs) and reverse
vector--Jacobian products (VJPs) for every differentiable primitive.
Differentiating energies, residuals, and matrix maps at converged states
realizes the stationary Lagrangian formulation of analytic-gradient
theory\cite{Handy_JChemPhys_1984_v81_p5031} while reusing the optimization
passes and CUDA execution pipeline of the primal equations.

\paragraph{Generation and execution.}
Source generation runs offline without loading the runtime library or probing a
GPU. Before emitting C++ and CUDA for an explicit hardware target, compiler
passes hoist iteration-invariant intermediates out of solver loops, fuse
elementwise operations, color scratch-buffer lifetimes into a symbolic memory
arena, and map tensor contractions onto vendor GEMM calls and specialized CUDA
kernels. For example, the closed-shell coupled-cluster and response paths alone
use 13 generator scripts to emit 34 C++ and CUDA source files containing 9,245
CUDA kernel definitions. Packaged builds compile the generated kernels ahead of
time (AOT) for the target architecture, whereas generated modules outside the
packaged profiles are compiled with NVCC on first use and cached. CPU and CUDA
backends share generated mathematical expressions, such as the VV10 pair
energies and derivatives, while maintaining separate execution schedules.

\paragraph{Generated and native components.}
The compiler generates the scientific programs: integral recurrences,
operator derivatives and contractions, most exchange--correlation point code,
the TensorIR method equations, and their precision variants and
operator-specific schedules. The native runtime provides the execution
infrastructure, including memory allocation, CUDA streams, events and graphs,
generic tiling and reductions, vendor libraries and the application
interfaces. Hand-written native code implements the iterative solvers,
dedicated integral, Fock-build, and derivative routines, the PBE and PBE0
point evaluator, and the VV10 pair traversal.

\subsection{Example: from the $\Lambda$ equations to DF-CCSD(T) forces}
\label{sec:methods-dfccsdt}

\begin{figure}[p]
  \centering
  \includegraphics[width=\textwidth]{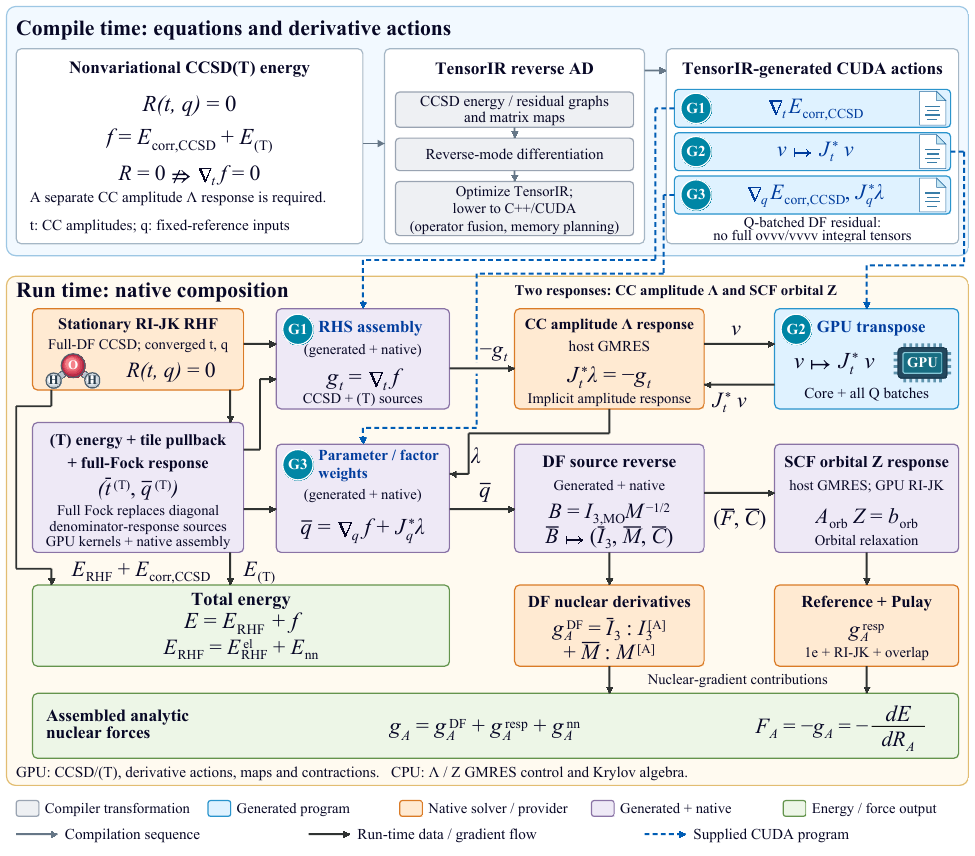}
  \caption{Example of the scientific compiler: DF-CCSD(T) energies and
    forces in the full-DF model. Energy and force calls share one converged
    state; the energy output adds
    $E_{\mathrm{RHF}}+E_{\mathrm{corr,CCSD}}$ from the RHF/DF-CCSD step and
    $E_{(T)}$ from the (T) module. At compile
    time, TensorIR reverse AD turns the CCSD energy and residual equations
    into the derivative programs G1--G3, and the run-time force driver uses
    each at the step with the same label; the host GMRES solver of the
    amplitude ($\Lambda$) response calls the transpose action G2 repeatedly.
    The native (T) module calls generated tile VJPs and supplies the sources
    $(\bar t^{(T)},\bar q^{(T)})$ with the full Fock replacement. Reversing the
    DF construction gives the integral weights for the DF nuclear derivatives
    and the orbital weights $\bar C$, which enter the separate
    self-consistent-field (SCF) orbital response $A_{\mathrm{orb}}Z=b_{\mathrm{orb}}$
    together with the Fock cotangent $\bar F$; compiler-generated matrix maps
    handle the matrix transformations in both steps. Symbols are defined in Section~\ref{sec:methods-dfccsdt}, those of
    the DF and gradient steps with eq~\ref{eq:methods-dfgrad}; $o$ and $v$ denote
    occupied and virtual orbitals, $Q$ the auxiliary functions, and 1e
    one-electron integral derivatives. In the correlation-only DF model, a
    conventional RHF reference replaces the RI-JK reference, and the orbital
    response and the reference terms use exact four-center integrals.}
  \label{fig:pipeline}
\end{figure}

Analytic CCSD(T) forces illustrate how the generated programs and the native
solvers share the work of a force calculation. Whereas a mean-field energy is
stationary with respect to the molecular
orbitals,\cite{Handy_JChemPhys_1984_v81_p5031} the CCSD(T) energy is stationary
neither in the cluster amplitudes nor in the reference orbitals; its analytic
gradient therefore requires the amplitude and orbital response in addition to
the derivatives of the energy
expression,\cite{Scuseria_JChemPhys_1991_v94_p442,Lee_JChemPhys_1991_v94_p6229}
and density fitting adds the derivatives of the three-center integrals and of
the auxiliary metric.\cite{Bozkaya_JChemPhys_2017_v147_p44104} In GenerativeQC, TensorIR AD builds the response programs of the DF-CCSD(T)
force chain from the CCSD equations, and native solvers and
integral-derivative routines use them in the complete force calculation of
Figure~\ref{fig:pipeline}.

Let $t=(t_1,t_2)$ be the converged CCSD amplitudes, $R(t,q)=0$ their
residuals, and $q$ the fixed-reference Fock, integral and DF-factor inputs.
Write $f(t,q)=E_{\mathrm{corr,CCSD}}(t,q)+E_{(T)}(t,q)$, so that the total
energy is $E=E_{\mathrm{RHF}}+f$, where $E_{\mathrm{RHF}}=E_{\mathrm{RHF}}^{\mathrm{el}}
+E_{\mathrm{nn}}$ is the electronic RHF energy plus the nuclear repulsion
$E_{\mathrm{nn}}$. In a force call, the (T) module returns $E_{(T)}$ together
with its sources $(\bar t^{(T)},\bar q^{(T)})$, the cotangents of $E_{(T)}$
with respect to $t$ and $q$. In these cotangents, the full occupied--occupied and virtual--virtual Fock
response replaces the diagonal orbital-energy terms so that the denominator
response is not double-counted. The converged
residuals do not make $f$ stationary in $t$, so the amplitude response needed
for the nuclear gradient is eliminated by solving the $\Lambda$ equations
\begin{equation}
  J_t^*\lambda=-g_t,\qquad
  g_t=\nabla_t f=\nabla_t E_{\mathrm{corr,CCSD}}+\bar t^{(T)},\qquad
  J_t=\frac{\partial R}{\partial t},
  \label{eq:methods-implicit}
\end{equation}
and then forming derivative weights for the electronic inputs,
\begin{equation}
  \bar q=\nabla_q f+J_q^*\lambda
  =\nabla_q E_{\mathrm{corr,CCSD}}+J_q^*\lambda+\bar q^{(T)},\qquad
  J_q=\frac{\partial R}{\partial q}.
  \label{eq:methods-parameter}
\end{equation}
The adjoints denoted by $*$ use the dense Frobenius inner product, restricted
to the amplitude symmetries. The native solver stores one representative of
each doubles pair $t_{ij}^{ab}=t_{ji}^{ba}$, weighted by $\sqrt{2}$ for a
pair of distinct entries and by one for a self-pair. This packing preserves
the inner product in the independent coordinates.

The compiler starts from the CCSD energy and residual graph. A reverse pass
on the energy generates the energy VJP (G1 in Figure~\ref{fig:pipeline}),
which gives $\nabla_t E_{\mathrm{corr,CCSD}}$; a pass seeded on the residual
generates the callable transpose action $J_t^*v$ (G2); and the parameter VJPs
(G3) give $\nabla_q E_{\mathrm{corr,CCSD}}$ and $J_q^*\lambda$. The native
driver solves eq~\ref{eq:methods-implicit} with
GMRES,\cite{Saad_SiamJSciStatComput_1986_v7_p856} which calls G2 repeatedly.
The DF transpose operator combines the retained CCSD core with every
auxiliary-index contribution to the virtual residual. In its staged form,
the core supplies seeds for shared intermediates, the native transpose action
calls the generated staged function once for each auxiliary batch to return
amplitude weights and intermediate seeds, and the accumulated seeds are
propagated through the shared amplitude map once. Retaining both direct
amplitude dependence and dependence through these intermediates completes
the chain rule without constructing the omitted three- and four-virtual
integral blocks.
After verifying the final residual with an independently expanded transpose
action, the force driver passes the resulting $\Lambda$ solution and its
parameter VJPs to the derivative assembly.

To reach nuclear derivatives, the runtime combines the retained-integral
and virtual-factor weights into the cotangent $\bar B$ of the DF factors and
reverses their construction $B=I_{3,\mathrm{MO}}M^{-1/2}$. Here
$I_{3,\mathrm{MO}}$ is the molecular-orbital transform, with coefficients
$C$, of the atomic-orbital three-center integrals $I_3$, $M$ is the
auxiliary Coulomb metric and $M^{-1/2}$ its retained fixed-rank inverse
square root. Generated matrix maps return
$\bar B\mapsto(\bar I_3,\bar M,\bar C)$; dedicated CUDA routines contract
these weights with the nuclear derivative integrals,
\begin{equation}
  g_A^{\mathrm{DF}}=\bar I_3\mathbin{:}I_3^{[A]}+\bar M\mathbin{:}M^{[A]},
  \label{eq:methods-dfgrad}
\end{equation}
where $X^{[A]}=\partial X/\partial R_A$ for nucleus $A$ and the colon
contracts all integral indices. The Fock cotangent $\bar F$ and $\bar C$
form the right-hand side $b_{\mathrm{orb}}$ of the orbital Z-vector
equation $A_{\mathrm{orb}}Z=b_{\mathrm{orb}}$, with orbital-response operator
$A_{\mathrm{orb}}$, whose solution gives the reference and
Pulay contribution $g_A^{\mathrm{resp}}$ from one-electron, RI-JK and overlap
derivatives. The driver then adds the nuclear-repulsion gradient
$g_A^{\mathrm{nn}}=\partial E_{\mathrm{nn}}/\partial R_A$ and negates the
total gradient to give
$F_A=-g_A=-(g_A^{\mathrm{DF}}+g_A^{\mathrm{resp}}+g_A^{\mathrm{nn}})=-dE/dR_A$.

Operator actions and tensor contractions run on the GPU, while the native
$\Lambda$ and orbital GMRES solvers keep their control flow and packed Krylov
vector algebra on the host. The solvers apply the orbital Hessian as an
operator, and the nuclear derivative integrals are contracted directly with
the weights, so neither is stored.

\subsection{Agent-driven performance optimization}
\label{sec:methods-optimization}

Generating correct code from symbolic equations does not by itself guarantee
high GPU performance: out-of-the-box language models rarely match the PyTorch
baseline in KernelBench without execution and profiling
feedback,\cite{Ouyang_arXiv_2025_p2502.10517} whereas evolutionary agents such
as AlphaEvolve discover faster kernels and algorithms by iterating against
automated evaluators.\cite{Novikov_arXiv_2025_p2506.13131}
In GenerativeQC, coding agents refine the compiler lowering rules and native
routines through an iterative loop of hypothesis generation, implementation,
evaluation and selection. Starting from timing breakdowns alongside iteration and
operator-evaluation counts, each candidate change tests a specific performance
hypothesis---reducing algebraic work, fusing operations and reusing resident
device buffers across phases, or tuning data layouts, schedules, and kernels.
Candidates are evaluated on complete molecular energy-and-force calls rather
than kernel microbenchmarks alone, because extra synchronization, data movement,
or memory-driven loop tiling can offset isolated kernel gains. In the
density-fitted response, for example, reusing resident J and K buffers avoids
recomputing auxiliary projections across tiled panels. Candidates are checked
against independent numerical references; those that shorten complete calls or
provide verified, mathematically equivalent simplifications of generated code
become defaults, whereas the rest are reverted or kept as explicit options.

%% file: sections/protocol.tex
\section{Benchmark results and discussion}
\label{sec:results}

The benchmarks compare the HF and DFT paths of GenerativeQC with GPU4PySCF and
evaluate the accuracy and computational cost of the CCSD(T)/CBS composite on
RTX~5090 and H100 GPUs.

\subsection{Computational setup and benchmark systems}
\label{sec:protocol}
\label{sec:protocol-systems}

For HF and DFT, each of the eight methods of Section~\ref{sec:methods-scope},
HF, PBE, PBE0, B3LYP, r$^2$SCAN, r$^2$SCAN-3c, M06-2X and $\omega$B97M-V, is
run on the Direct and DF paths for water clusters of 3, 12, 48 and 96 atoms
(1, 4, 16 and 32 molecules). The clusters are nested subsets of one
synthetic 96-atom structure, a $2\times2$ array of copies of the S4 water
octamer of the WATER27
set\cite{Bryantsev_JChemTheoryComput_2009_v5_p1016,Goerigk_PhysChemChemPhys_2017_v19_p32184}
translated by $\pm\SI{5}{\angstrom}$ along two Cartesian axes, forming a size
series of neutral closed-shell singlets.
The orbital basis is spherical
def2-SVP\cite{Weigend_PhysChemChemPhys_2005_v7_p3297} (24 to 768
basis functions), except that r$^2$SCAN-3c uses its def2-mTZVPP basis with D4
and gCP corrections, which enter both the timings and the compared results.
These 64 method--path--size cases are measured on each GPU using the default
convergence and grid settings of GenerativeQC and GPU4PySCF~1.9.0, with
complete grid-response forces enabled in both programs (SI
Section~S1).\cite{Li_JPhysChemA_2025_v129_p1459,Wu_WiresComputMolSci_2025_v15_pe70008}
Both programs stop the SCF iterations when the energy change falls below
$10^{-9}~E_\mathrm{h}$ and the occupied--virtual orbital-gradient norm falls
below $\sqrt{10^{-9}}$, while retaining their respective standard defaults for
the initial guess and for integral and exchange--correlation screening (SI
Table~S1).

For CCSD(T)/CBS we use the seven $n$-alkanes C$_n$H$_{2n+2}$ ($n=1$--7,
denoted C1--C7, 5 to 23 atoms) of the QM7 data
set,\cite{Blum_JAmChemSoc_2009_v131_p8732,Rupp_PhysRevLett_2012_v108_p058301}
with zero-based indices 0, 1, 8, 48, 183, 293 and 5604, at their stored
geometries. The composite is the all-electron full-DF scheme of
Section~\ref{sec:methods-scope}: cc-pVDZ/cc-pVTZ extrapolation and a cc-pVDZ
correction with the matching JKFIT auxiliaries. The orbital basis grows from
34/86 (C1) to 178/434 (C7) functions and the auxiliary basis from 162/199 to
858/1033 functions in the cc-pVDZ/cc-pVTZ groups.

All timings are complete energy-and-force calls on a single RTX~5090 or H100
GPU with one math thread. GenerativeQC is built with CUDA~12.9 and GPU4PySCF
runs on PySCF~2.14.0,\cite{Sun_JChemPhys_2020_v153_p24109} with full software
environment details given in SI Section~S1. For HF and DFT, the 64 pairs are measured in three passes (the second in
reverse order), with the two programs running consecutively in fresh processes
on the same GPU and each running first in half of the pairs. Each process
executes one cold and three warm calls (1536 timed calls per GPU); the cold call
includes object construction and preparation with prebuilt code and excludes
process start-up and compilation. Each warm call repeats the calculation at
the same geometry from the process's own frozen cold density. For each case,
we report the median and range over the three processes for the cold call and
for the per-process median of the three warm calls. For
CCSD(T)/CBS, each reported time is the median of two calls using a prebuilt
binary, where each call constructs a new calculator and
returns the complete energy and forces, excluding process and CUDA start-up and
the preparation of basis-set data.

The Supporting Information reports the errors of the HF and DFT energies and
forces against tightly converged, denser-grid GPU4PySCF references checked by
3-atom PySCF calculations (SI Section~S2), as well as the convergence
controls, independent PySCF validation and timings of the CCSD(T)/CBS composite
(SI Section~S3).

%% file: sections/results.tex
\subsection{HF and DFT energies and forces}
\label{sec:results-hfdft}

\begin{figure*}
  \centering
  \includegraphics[width=\textwidth]{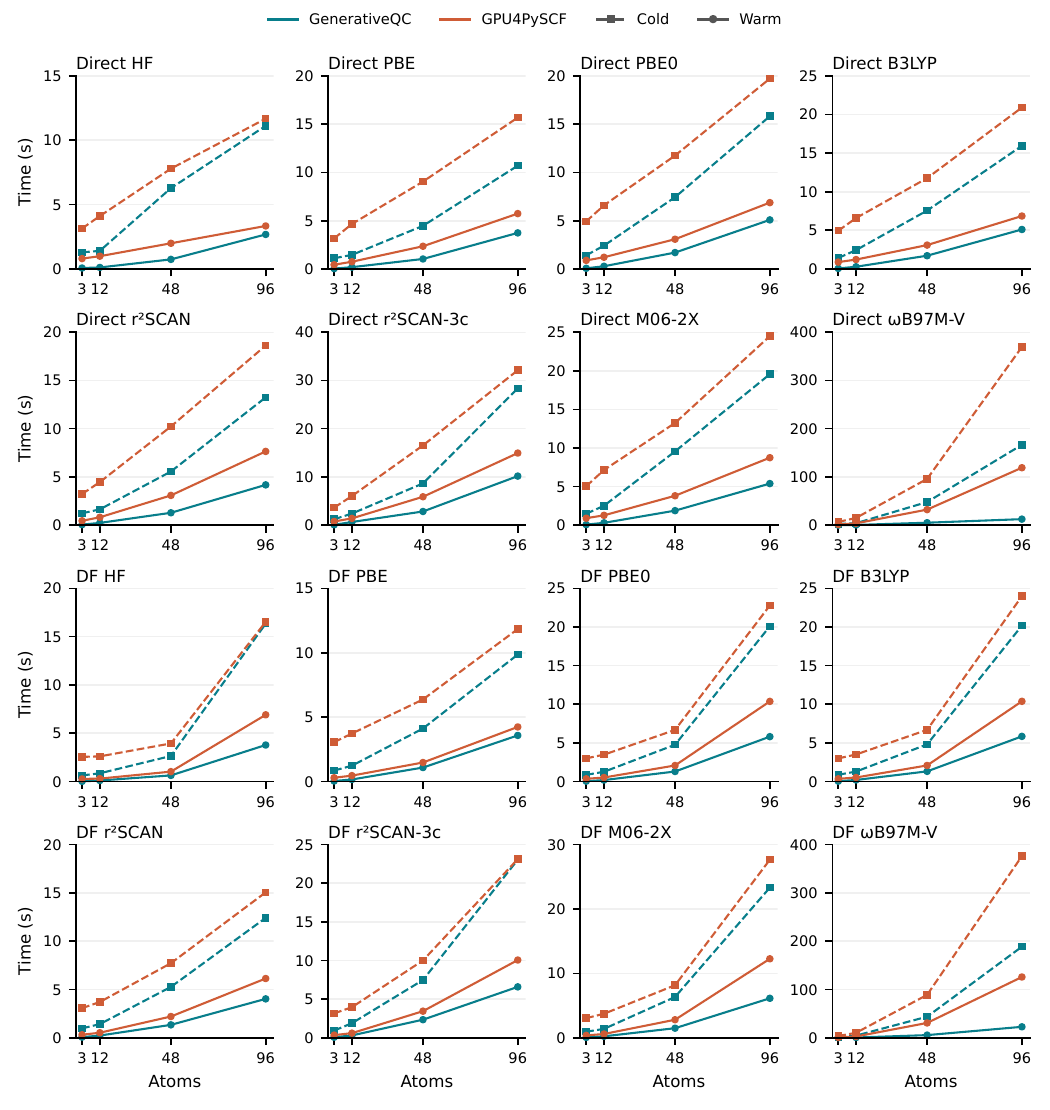}
  \caption{Complete energy-and-force times of GenerativeQC and GPU4PySCF~1.9.0
    on RTX~5090 for water clusters of 3, 12, 48 and 96 atoms, on the Direct
    (upper two rows) and DF (lower two rows) paths; both axes are linear, and
    the time axis is scaled separately in each panel.
    Dashed lines with squares show the cold call and solid lines with circles
    show the per-process warm-call median, each plotted as the median over
    three processes. The basis is def2-SVP, except def2-mTZVPP with D4 and gCP
    for r$^2$SCAN-3c. SI Table~S3 lists the medians and three-process ranges,
    and SI Figure~S1 shows the medians on logarithmic axes.}
  \label{fig:hfdft-rtx5090}
\end{figure*}

\begin{figure*}
  \centering
  \includegraphics[width=\textwidth]{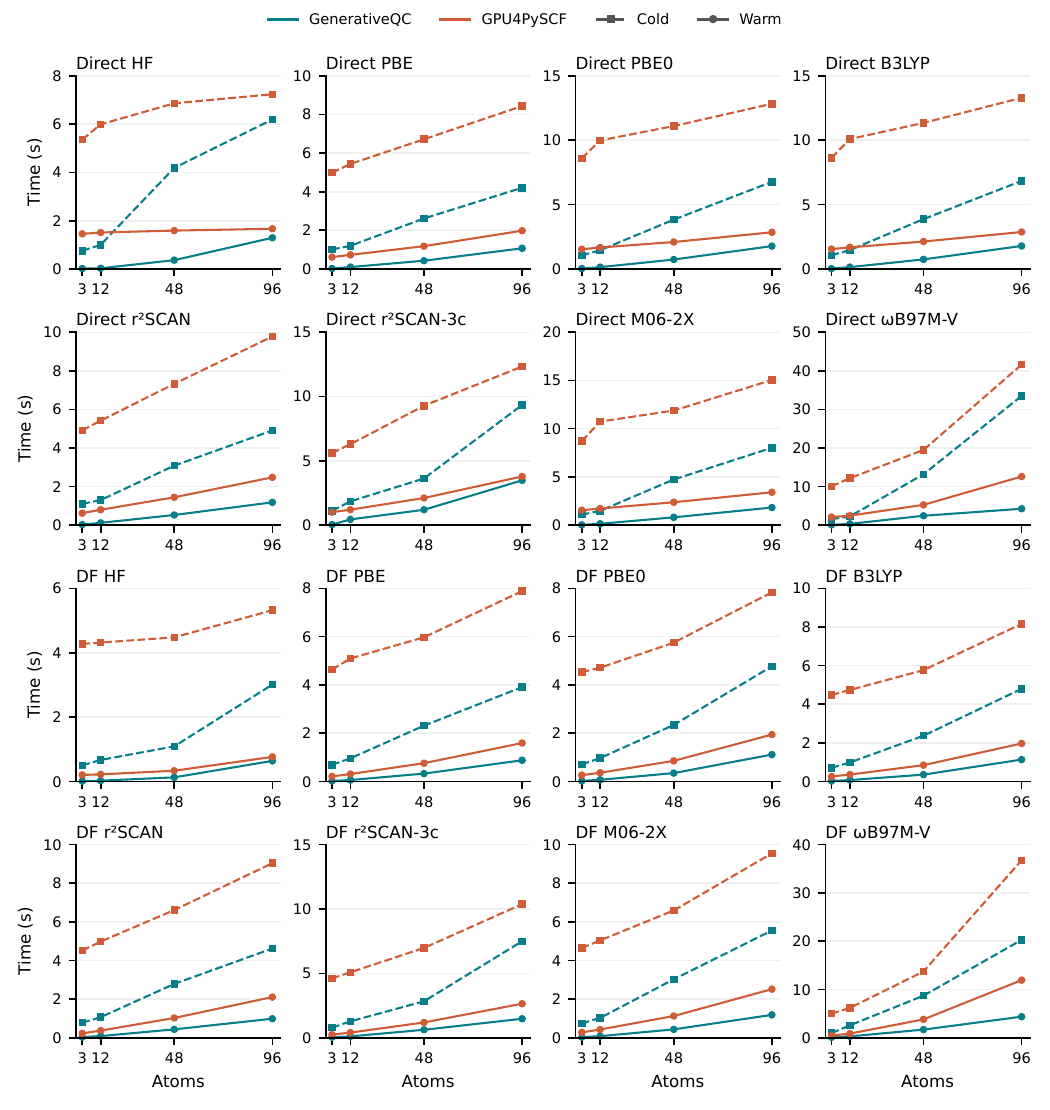}
  \caption{As Figure~\ref{fig:hfdft-rtx5090}, on H100. SI Table~S4 lists the
    medians and three-process ranges, and SI Figure~S2 shows the medians on
    logarithmic axes.}
  \label{fig:hfdft-h100}
\end{figure*}

GenerativeQC computes HF and DFT energies and forces faster than GPU4PySCF in
all 64 cold and 64 warm comparisons on both RTX~5090 and H100 GPUs
(Figures~\ref{fig:hfdft-rtx5090} and~\ref{fig:hfdft-h100}). Summed across all
64 cases, the median cold-call time totals
\GQC{hfdft.rtx5090.total.cold.gqc}~s versus
\GQC{hfdft.rtx5090.total.cold.gpu4pyscf}~s on RTX~5090 and
\GQC{hfdft.h100.total.cold.gqc}~s versus
\GQC{hfdft.h100.total.cold.gpu4pyscf}~s on H100, while the median warm-call
time totals \GQC{hfdft.rtx5090.total.warm.gqc}~s versus
\GQC{hfdft.rtx5090.total.warm.gpu4pyscf}~s on RTX~5090 and
\GQC{hfdft.h100.total.warm.gqc}~s versus
\GQC{hfdft.h100.total.warm.gpu4pyscf}~s on H100.

Relative speedups are largest for small clusters. GenerativeQC executes a
prepared calculation plan natively in C++ and CUDA without per-iteration Python
dispatch, and warm calls verify convergence in a single Fock or
effective-potential build (two for DF HF), compared with two builds in every
GPU4PySCF warm call. At 3 atoms, a warm GenerativeQC call takes a median of
\GQC{hfdft.rtx5090.small.warm.gqc.ms}~ms on RTX~5090 and
\GQC{hfdft.h100.small.warm.gqc.ms}~ms on H100, compared with
\GQC{hfdft.rtx5090.small.warm.gpu4pyscf.ms} and
\GQC{hfdft.h100.small.warm.gpu4pyscf.ms}~ms for GPU4PySCF, yielding median
speedups of $\GQC{hfdft.rtx5090.warm.median.3}\times$ and
$\GQC{hfdft.h100.warm.median.3}\times$
($\GQC{hfdft.rtx5090.cold.median.3}\times$ and
$\GQC{hfdft.h100.cold.median.3}\times$ for cold calls). Even for cold calls,
every 3- and 12-atom case is at least twice as fast on both GPUs.

As cluster size grows, the absolute time saved per warm call increases on both
GPUs, from a median of \GQC{hfdft.rtx5090.saved.warm.3}~s at 3 atoms to
\GQC{hfdft.rtx5090.saved.warm.96}~s at 96 atoms on RTX~5090 and from
\GQC{hfdft.h100.saved.warm.3} to \GQC{hfdft.h100.saved.warm.96}~s on H100. At 96
atoms, warm GenerativeQC calls finish in
\GQC{hfdft.rtx5090.large.warm.gqc.min}--\GQC{hfdft.rtx5090.large.warm.gqc.max}~s
on RTX~5090 and
\GQC{hfdft.h100.large.warm.gqc.min}--\GQC{hfdft.h100.large.warm.gqc.max}~s on
H100, maintaining median speedups of
$\GQC{hfdft.rtx5090.warm.median.96}\times$ and
$\GQC{hfdft.h100.warm.median.96}\times$
($\GQC{hfdft.rtx5090.cold.median.96}\times$ and
$\GQC{hfdft.h100.cold.median.96}\times$ for cold calls). The largest absolute
savings occur for the most demanding functional, $\omega$B97M-V. At 48 and 96
atoms, warm $\omega$B97M-V calls are
$\GQC{hfdft.rtx5090.wbmv.min}\text{--}\GQC{hfdft.rtx5090.wbmv.max}\times$
faster than GPU4PySCF on RTX~5090 (where GPU4PySCF slows down much more sharply
than GenerativeQC) and
$\GQC{hfdft.h100.wbmv.min}\text{--}\GQC{hfdft.h100.wbmv.max}\times$ faster on
H100, reducing a 96-atom \GQC{hfdft.rtx5090.wbmv96.path} $\omega$B97M-V warm
call from \GQC{hfdft.rtx5090.wbmv96.gpu4pyscf} to
\GQC{hfdft.rtx5090.wbmv96.gqc}~s on RTX~5090 and from
\GQC{hfdft.h100.wbmv96.gpu4pyscf} to \GQC{hfdft.h100.wbmv96.gqc}~s on H100. In
these warm calls, both programs converge in one SCF iteration, and GenerativeQC
reuses stored VV10 energy and potential coefficients when the per-point inputs
match bitwise (SI Section~S1).

Across all 64 cases on both GPUs, matching GenerativeQC and GPU4PySCF calls
agree to within $\GQC{hfdft.directdiff.energy}~E_\mathrm{h}$ per atom in energy
and $\GQC{hfdft.directdiff.force}~E_\mathrm{h}/a_0$ in every force component;
against tightly converged, denser-grid references, both programs exhibit the
same density-fitting and grid-quadrature errors (SI Section~S2 and Table~S2).

\subsection{CCSD(T)/CBS energies and forces}
\label{sec:results-cbs-cost}

\begin{figure*}
  \centering
  \includegraphics[width=\textwidth]{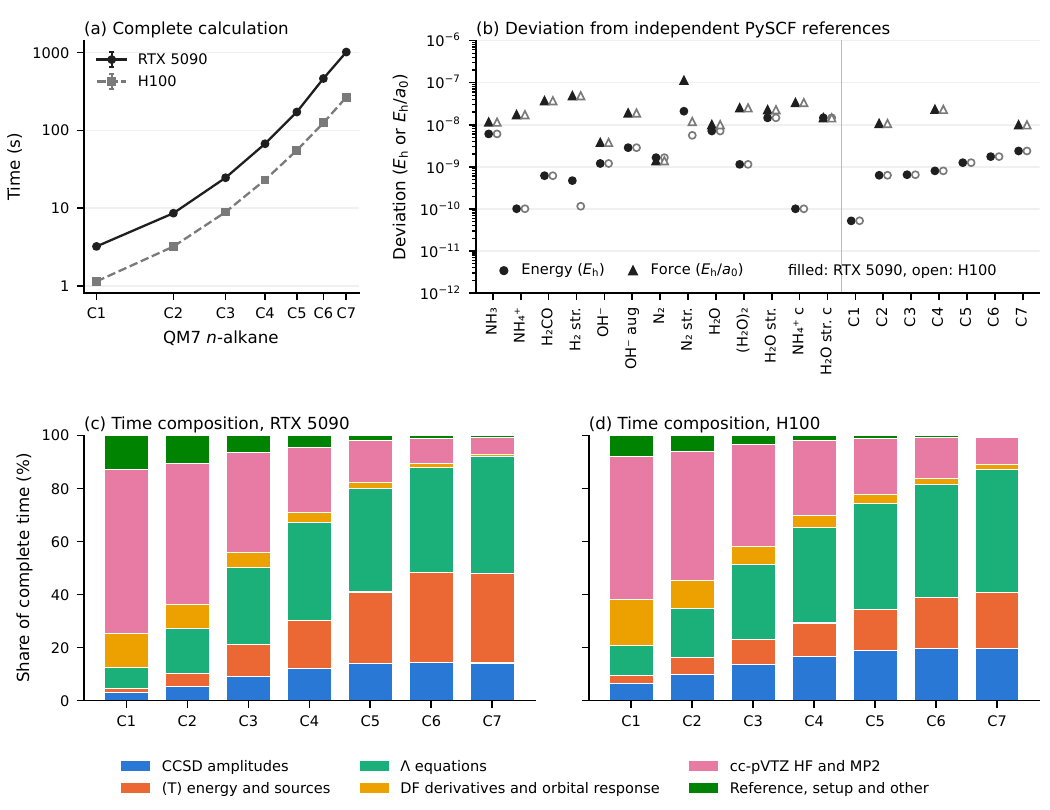}
  \caption{Default full-DF CCSD(T)/CBS composite (cc-pVDZ/cc-pVTZ MP2
    extrapolation plus the cc-pVDZ DF-CCSD(T)$-$DF-MP2 correction) on RTX~5090
    and H100. (a) Complete energy-and-force time for the QM7 $n$-alkanes C1--C7,
    plotted against the number of atoms on logarithmic axes; median of two calls
    with their range. (b) Deviations from independent PySCF references, in
    $E_\mathrm{h}$ for energies and $E_\mathrm{h}/a_0$ for forces: the largest
    energy and force-component errors of 13 small-molecule test cases (c,
    correlation-only DF model; str., stretched; aug, augmented basis), and the
    composite energies of C1--C7 and directional force projections of C2, C4
    and C7. Filled and open markers show RTX~5090 and H100. (c, d) Share of the
    complete time spent in each step, mean of the two calls.}
  \label{fig:cbs}
\end{figure*}

GenerativeQC evaluates complete CCSD(T)/CBS energies and analytic forces for
the QM7 $n$-alkanes from methane (C1, 5 atoms) to heptane (C7, 23 atoms) on a
single GPU (Figure~\ref{fig:cbs}a). Total runtimes grow from
\GQC{cbs.rtx5090.first}~s on RTX~5090 and \GQC{cbs.h100.first}~s on H100 for
methane to \GQC{cbs.rtx5090.last.min}~min (\GQC{cbs.rtx5090.last}~s) on
RTX~5090 and \GQC{cbs.h100.last.min}~min (\GQC{cbs.h100.last}~s) on H100 for
heptane. As the carbon chain lengthens, the cc-pVDZ group (HF, MP2, and
DF-CCSD(T)) increasingly dominates the runtime over the cc-pVTZ extrapolation
group, reaching \GQC{cbs.rtx5090.share.last}\% for heptane on RTX~5090 and
\GQC{cbs.h100.share.last}\% on H100, with the perturbative (T) step and
$\Lambda$ equations alone accounting for \GQC{cbs.rtx5090.tlambda.last}\% and
\GQC{cbs.h100.tlambda.last}\% (Figure~\ref{fig:cbs}c,d; SI Table~S8).

Across all seven alkanes, the composite energies agree with independent PySCF
calculations to within $\GQC{cbs.alkane.energy.bound}~E_\mathrm{h}$, and
directional force projections for ethane, butane, and heptane agree with
independent finite differences to within
$\GQC{cbs.alkane.direction.bound}~E_\mathrm{h}/a_0$ (Figure~\ref{fig:cbs}b).
Across 13 additional small-molecule test cases with complete Cartesian reference
forces, the largest deviations are $\GQC{cbs.val13.energy.bound}~E_\mathrm{h}$
in energy and $\GQC{cbs.val13.force.bound}~E_\mathrm{h}/a_0$ in a force
component (SI Section~S3). Relative to the tight convergence controls of SI
Table~S5, the default controls shorten the heptane calculation by
\GQC{cbs.rtx5090.saving.last}\% on RTX~5090 and \GQC{cbs.h100.saving.last}\% on
H100 while leaving the alkane energies and forces unchanged to within
$\GQC{cbs.strict.energy.bound}~E_\mathrm{h}$ and
$\GQC{cbs.strict.force.bound}~E_\mathrm{h}/a_0$.

%% file: sections/conclusion.tex
\section{Conclusions}
\label{sec:conclusion}

We have presented GenerativeQC, an open-source, GPU-accelerated quantum
chemistry program built around a scientific compiler and a native C++/CUDA
runtime. By translating symbolic method equations and applying compile-time
automatic differentiation to derive derivative and response programs,
GenerativeQC computes energies and analytic nuclear forces from HF and DFT to
MP2, DF-CCSD(T), and an MP2-assisted CCSD(T)/\allowbreak{}CBS composite
alongside hand-written solvers and integral routines.
An iterative, agent-driven optimization loop further refines the compiler
lowering rules and native routines, promoting changes to defaults when they
maintain agreement with independent references and shorten complete
energy-and-force calls.

In the HF and DFT benchmarks, GenerativeQC completes energy-and-force
calculations faster than GPU4PySCF in every cold and warm comparison on both
RTX~5090 and H100 GPUs, with the largest relative speedups on the smallest
clusters. Across all 64 cases on both GPUs, matching calls of the two programs
agree to within $\GQC{hfdft.directdiff.energy}~E_\mathrm{h}$ per atom in energy
and $\GQC{hfdft.directdiff.force}~E_\mathrm{h}/a_0$ in every force component.

For correlated methods, GenerativeQC evaluates complete analytic forces for
closed-shell DF-CCSD(T) and the MP2-assisted CCSD(T)/CBS composite on GPUs,
executing compiler-generated derivative contractions and transpose Jacobian
operators for the (T) contribution and the amplitude and orbital responses. In 13
small-molecule test cases and checks on the QM7 alkanes, the composite
energies and forces agree with independent PySCF calculations to within
$\GQC{cbs.all.energy.bound}~E_\mathrm{h}$ and
$\GQC{cbs.all.force.bound}~E_\mathrm{h}/a_0$ on both RTX~5090 and H100 GPUs,
and a complete
energy-and-force calculation for heptane takes \GQC{cbs.rtx5090.last.min}~min
on RTX~5090 and \GQC{cbs.h100.last.min}~min on H100.

%% file: sections/availability.tex
\section*{Data and software availability}

\begin{sloppypar}
GenerativeQC is publicly
available at \url{https://github.com/jinzhezenggroup/generativeqc}.
\end{sloppypar}

\begin{acknowledgement}
J.Z. was supported by the Advanced Materials National Science and Technology
Major Project (No.~2025ZD0618702), the National Natural Science Foundation of
China (Grant No.~22603140), and the Fundamental Research Funds for the Central Universities.
T.Z. was supported by the National Natural Science
Foundation of China (Grant Nos.~92461313 and 22361132538), the AI for Science
Program of the Shanghai Municipal Commission of Economy and Informatization
(Grant No.~2025-GZL-RGZN-BTBX-01004), and the Open Research Fund of Suzhou
Laboratory (No.~SZLAB-1508-2024-TS017).
The AI-driven experiments, simulations, and model training were performed on the robotic AI-Scientist platform of the Chinese Academy of Sciences.
The numerical calculations were performed on the supercomputing system in the Supercomputing Center of the University of Science and Technology of China.
\end{acknowledgement}

\begin{suppinfo}
HF/DFT comparison settings (Table~S1), errors against the references
(Table~S2) and all timings on RTX~5090 and H100 (Tables~S3 and~S4, and
Figures~S1 and~S2 on logarithmic axes); CCSD(T)/CBS convergence
controls (Table~S5), validation errors for 13 small-molecule test cases and the
QM7 alkanes (Tables~S6 and~S7), and timings with their composition
(Table~S8).
\end{suppinfo}